\documentclass[fleqn,usenatbib]{mnras}
\usepackage{newtxtext,newtxmath}
\usepackage[T1]{fontenc}
\usepackage{ae,aecompl}

\usepackage{graphicx}	
\usepackage{amsmath}	
\usepackage{bm}
\usepackage{mathtools}
\usepackage{todonotes}
\usepackage{float}
\usepackage{array, makecell}
\usepackage{booktabs}
\usepackage{xcolor}

\newcommand{\markup}[1]{{#1}}

\renewcommand{\vec}[1]{{\bm{#1}}}
\newcommand{\oper}[1]{{\bm{\mathsf{#1}}}}
\newcommand{\software}[1]{{\textsc{#1}}}

\newcommand{\noisecov}[0]{\oper{C}^{-1}}
\newcommand{\imcov}[0]{\tilde{\oper{C}}_x^{-1}}
\newcommand{\imcovexplicit}[0]{\oper{C}_x^{-1}}
\newcommand{\dft}[0]{\oper{D}}
\newcommand{\nufft}[0]{\tilde{\dft}}
\newcommand{\fft}[0]{\oper{F}}
\newcommand{\lensop}[0]{\oper{L}}
\newcommand{\regul}[0]{\oper{R}}
\newcommand{\gridder}[0]{\oper{G}}
\newcommand{\zpad}[0]{\oper{Z}}
\newcommand{\apod}[0]{\oper{W}}
\newcommand{\msol}[0]{\oper{A}}
\newcommand{\dbconv}[0]{\oper{B}}
\newcommand{\pc}[0]{\oper{P}}
\newcommand{\source}[0]{\vec{s}}
\newcommand{\smp}[0]{\source_\mathrm{MP}}
\newcommand{\lams}[0]{\lambda_{\source}}
\newcommand{\data}[0]{\vec{d}}
\newcommand{\noise}[0]{\vec{n}}
\newcommand{\etalens}[0]{\vec{\eta}}
\newcommand{\nvis}[0]{N_\mathrm{vis}}
\newcommand{\ngrid}[0]{N_\mathrm{grid}}
\newcommand{\nsource}[0]{N_\mathrm{src}}
\newcommand{\logdet}[0]{\log \det}
\newcommand{\wsup}[0]{w_\mathrm{sup}}

\newcommand{\sgt}[0]{\source_\mathrm{GT}}
\newcommand{\etagt}[0]{\etalens_\mathrm{GT}}

\newcommand{\uu}[0]{\vec{u}}
\newcommand{\xx}[0]{\vec{x}}
\newcommand{\kk}[0]{\vec{k}}
\newcommand{\bb}[0]{\vec{b}}

\newcommand{\lamfid}[0]{\lambda_{\source,\mathrm{fid}}}
\newcommand{\snrfid}[0]{\mathrm{SNR}_\mathrm{fid}}

\newcommand{\comments}[1]{}

\newcommand{\revisions}[1]{#1}

\graphicspath{figures}

\title[\markup{Visibility-space gravitational lens modelling}]{A novel approach to visibility-space modelling of interferometric gravitational lens observations at high angular resolution}

\author[D. Powell et al.]{
Devon Powell$^{1}$\thanks{E-mail: dmpowell@mpa-garching.mpg.de},
Simona Vegetti$^{1}$,
John P. McKean$^{2,3}$,
Cristiana Spingola$^{4,5}$,
\newauthor
\revisions{Francesca Rizzo$^{1}$, and
Hannah R. Stacey$^{1,2,3}$}
\\
$^{1}$Max Planck Institute for Astrophysics, Karl-Schwarzschild-Stra\ss{}e 1, 85748 Garching bei M\"unchen, Germany\\
$^{2}$Kapteyn Astronomical Institute, University of Groningen, PO Box 800, NL-9700 AV Groningen, The Netherlands\\
$^{3}$ASTRON, Netherlands Institute for Radio Astronomy, PO Box 2, NL-7990 AA Dwingeloo, The Netherlands\\
$^{4}$INAF $-$ Istituto di Radioastronomia, via Gobetti 101, I$-$40129, Bologna, Italy \\
$^{5}$Dipartimento di Fisica e Astronomia, Universit\`a degli Studi di Bologna, via Gobetti 93/2, I$-$40129 Bologna, Italy \\
}

\date{Accepted 2020 August 29. Received 2020 August 4; in original form 2020 May 7.}

\pubyear{2020}

\begin{document}
\label{firstpage}
\pagerange{\pageref{firstpage}--\pageref{lastpage}}
\maketitle

\begin{abstract}
We present a new gravitational lens modelling technique designed to model high-resolution interferometric observations with large numbers of visibilities without the need to pre-average the data in time or frequency. We demonstrate the accuracy of the method using validation tests on mock observations. Using small data sets with $\sim 10^3$ visibilities, we first compare our approach with the more traditional direct Fourier transform (DFT) implementation and direct linear solver. Our tests indicate that our source inversion is indistinguishable from that of the DFT. Our method also infers lens parameters to within 1 to 2 per cent of both the ground truth and DFT, given sufficiently high signal-to-noise ratio (SNR). \revisions{ When the SNR is as low as 5, both approaches lead to errors of several tens of per cent in the lens parameters and a severely disrupted source structure, indicating that this is related to the SNR and choice of priors rather than the modelling technique itself.} We then analyze a large data set with $\sim 10^8$ visibilities and a SNR matching real global Very Long Baseline Interferometry observations of the gravitational lens system MG J0751+2716. The size of the data is such that it cannot be modelled with traditional implementations. Using our novel technique, we find that we can infer the lens parameters and the source brightness distribution, respectively, with an RMS error of $0.25$ and $0.97$ per cent relative to the ground truth.
\end{abstract}

\begin{keywords}
methods: data analysis -- techniques: high angular resolution -- techniques:image processing -- gravitational lensing: strong  
\end{keywords}


\section{Introduction}

Strong gravitational lensing by galactic-scale potentials is a powerful tool in astronomy, providing several routes towards independent constraints for astrophysical and cosmological models (see \citealt{treu2010} for a review on the subject).  

The most obvious feature is the magnification introduced by the lens, which drastically increases the effective angular resolution of the observations.  This has been taken advantage of by many authors \citep[e.g.][]{swinbank2015, lee2016,johnson2017, frizzo2018, spingola2019}, who leveraged the lensing effect to make detailed observations of galaxies at high redshifts that would be otherwise impossible to probe using current instrumentation. 

The differing lines of sight taken by multiple lensed images can be used to break degeneracies between properties of the source and lens. For instance, \citet{henkel2005}, \citet{marshall2017} and \citet{allison2017} use this effect to study spectral line absorption in the foreground lens galaxy. Similarly, \citet{mao2017} have recently studied plasma effects in gravitationally-lensed quasars in order to quantify polarization properties of the background source and magnetic fields in the lens galaxy. 

Gravitational lensing can also reveal the presence of low-mass dark matter haloes via their gravitational effect on the lensed images, and therefore, place constraints on the physical properties of dark matter \citep[e.g.][]{vegetti2009,vegetti2010,vegetti2012,vegetti2014a, hezaveh2016b,hezaveh2016a,birrer2017,gilman2019,ritondale2019,hsueh2019}. In combination with other measurements, it can be successfully used to constrain the structure of the lens galaxies, and therefore the processes that regulate the evolution of these objects \citep[e.g.][]{Koopmans09,Auger10a,Auger10b}.

It is, of course, advantageous to model gravitational lenses at high angular resolution and sensitivity in order to extract as much information as possible from each observation. In modern astronomy, this capability is provided by radio interferometers, which construct a synthetic aperture from collections of widely separated antennas to achieve an angular resolution not possible with optical telescopes \citep{smirnov2011}. While the idea of applying aperture synthesis techniques at radio frequencies is not new \citep[e.g.][]{ryle1950, jennison1958}, modern technological advances have since brought the field into maturity and now can provide observations at milli-arcsec-level (mas) resolution and high signal-to-noise ratio (SNR). These characteristics, make observations using Very Long Baseline Interferometry (VLBI) ideally suited to search for low-mass dark matter haloes \citep{mckean2015} or to study super-massive black holes associated with active galactic nuclei (AGN) in the early Universe \citep{spingola2019b} with strong gravitational lensing. 

However, there are subtleties inherent in the modelling of radio data that must be properly treated in order to obtain robust results. Namely, in contrast to optical and infrared telescopes, which observe the sky directly, interferometers operating at m to mm-wavelengths observe Fourier components of the sky that must be transformed back into image space. The challenge here is that the Fourier plane of the sky (the $uv$ plane; see Section \ref{sec:bayes}) is incompletely sampled, meaning that the transformation into image space is an ill-posed problem. Synthesizing images from radio data (deconvolution) to obtain a robust approximation for the sky surface brightness distribution is a well-studied topic with a large body of literature \citep[e.g.][]{hogbom1974,pearson1984,cornwell1985,sault1994,rau2011,wsclean2014,junk2016}. 

The challenges of modelling radio interferometric data sets are compounded when one considers that in gravitational lens modelling, the sky itself is a distorted and magnified version of a distant source. Therefore, one must propagate uncertainties to the source model in a self-consistent way. Na\"{i}vely, one might attempt to first image the sky using an established deconvolution technique, then apply a lens modelling code to this image. However, it is unclear how residual noise and other imaging artifacts would then propagate to the source and lens model. Moreover, pure deconvolution algorithms do not guarantee the conservation of source surface brightness, which is a fundamental feature of gravitational lensing. That is, the surface brightnesses of multiple lensed images of the same source may not be consistent with one another, which poses an obvious problem for the source reconstruction. 

We aim instead to fit a source and lens model to the data self-consistently in visibility space, an approach that has found some prior interest. For instance, \cite{Kochanek92} have introduced \software{lensclean}, a lens modelling method for interferometric data based on the \software{clean} algorithm. It was later on improved upon by \cite{Ellithorpe96} and \cite{Wucknitz04}. More recently, \cite{bussman2012,bussman2013} introduce a simple $\chi^2$ fit using a gradient descent method and posterior sampling to data from the Sub-Millimetre Array (SMA), using a parametric description of the background source. \cite{rybak2015b} and \cite{hezaveh2016b} use a Gaussian likelihood and prior to form a linear least-squares equation in order to analyze data from the Atacama Large Millimetre Array (ALMA) with pixellated sources. However, these approaches were highly limited by the size of the data. Therefore, they rely heavily on averaging the data to a manageable size before the fitting is carried out. As time or frequency averaging smears the visibiliies in the $uv$ plane tangentially and radially, respectively, it corrupts the data leading to a loss of spatial information. In particular, such averaging results in a lowering of the sky surface brightness emission as a function of distance from the correlated delay centre. This is especially critical for VLBI observations of fields that are even just a few arcsec in size, where little or no time or frequency averaging can be achieved without smearing.

In this paper, we present an advanced Bayesian code for gravitational lens modelling that directly fits radio interferometric data sets in visibility space, with no need for averaging or otherwise reducing the data size beforehand. We demonstrate that it is possible to directly model VLBI data, which can contain large numbers ($>10^8$) of visibilities. This work is an extension of the framework developed by \cite{vegetti2009}, \cite{rybak2015b}, and \cite{frizzo2018}. The modification is mathematically straightforward, but presents a unique set of computational challenges that we overcome.

We first give a brief introduction to the Bayesian inference process (Section \ref{sec:bayes}) for mathematical context.  We then describe in detail our numerical methods for directly fitting gravitational lens models to large radio visibility data sets that until now have been intractable (Section \ref{sec:radioimager}). In Section \ref{sec:simobs} we describe several tests on mock observations to show that our implementation of the algorithm performs robustly. These mock obervations were created starting from a model of real global VLBI observations of the lens system MG J0751+2716, presented in Section \ref{sec:modelling}. We summarise our main findings in Section \ref{sec:conclusions}.

\section{Hierarchical Bayesian inference on interferometric data} \label{sec:bayes}

 \begin{figure*}
 \centering
  \includegraphics[scale=0.45]{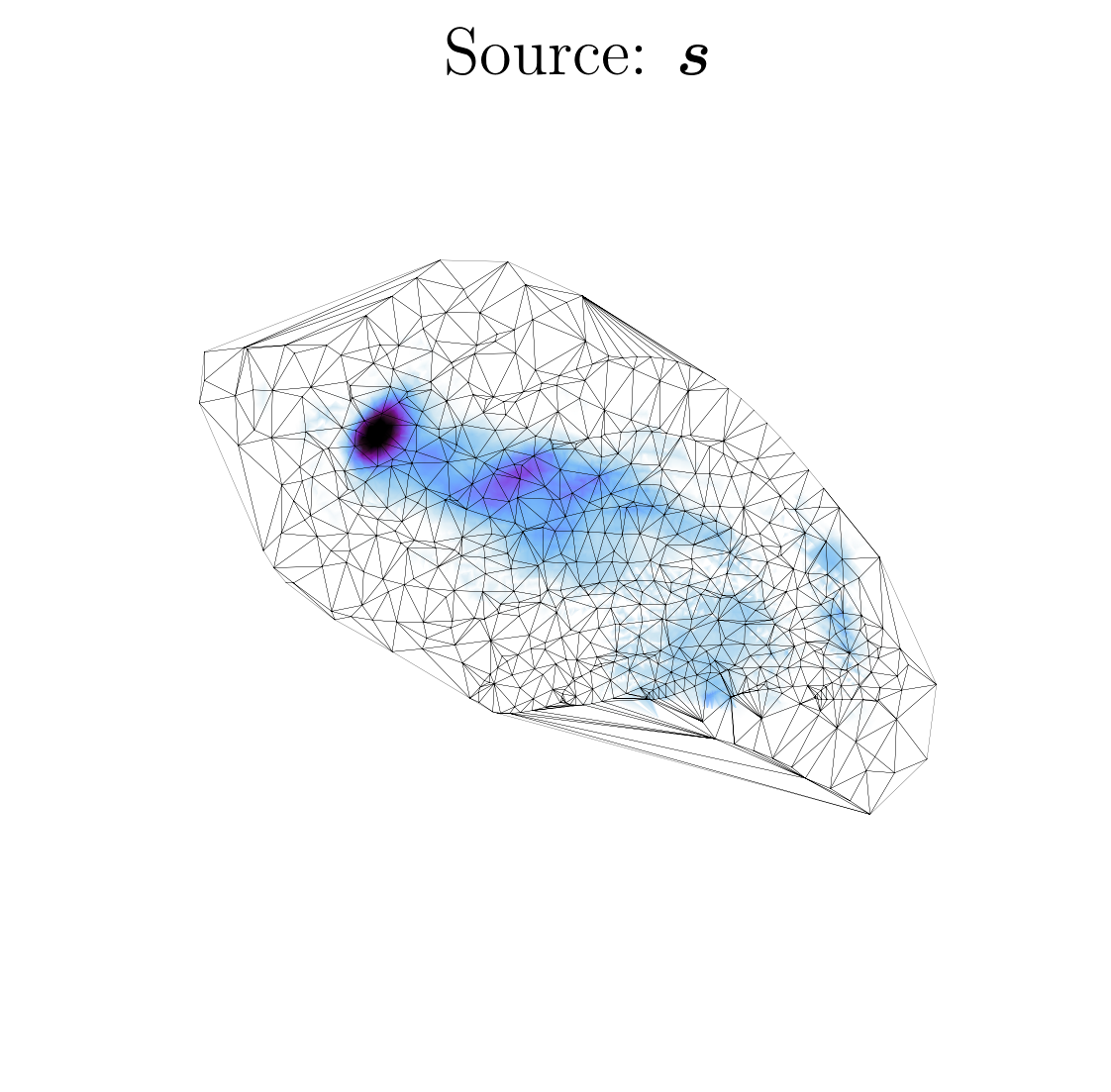}
  \includegraphics[scale=0.45]{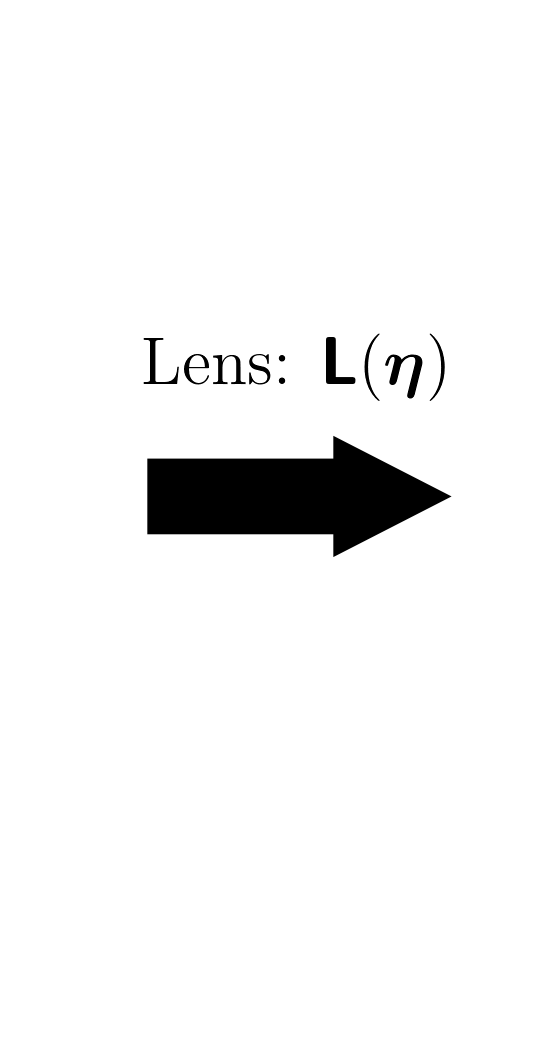}
  \includegraphics[scale=0.45]{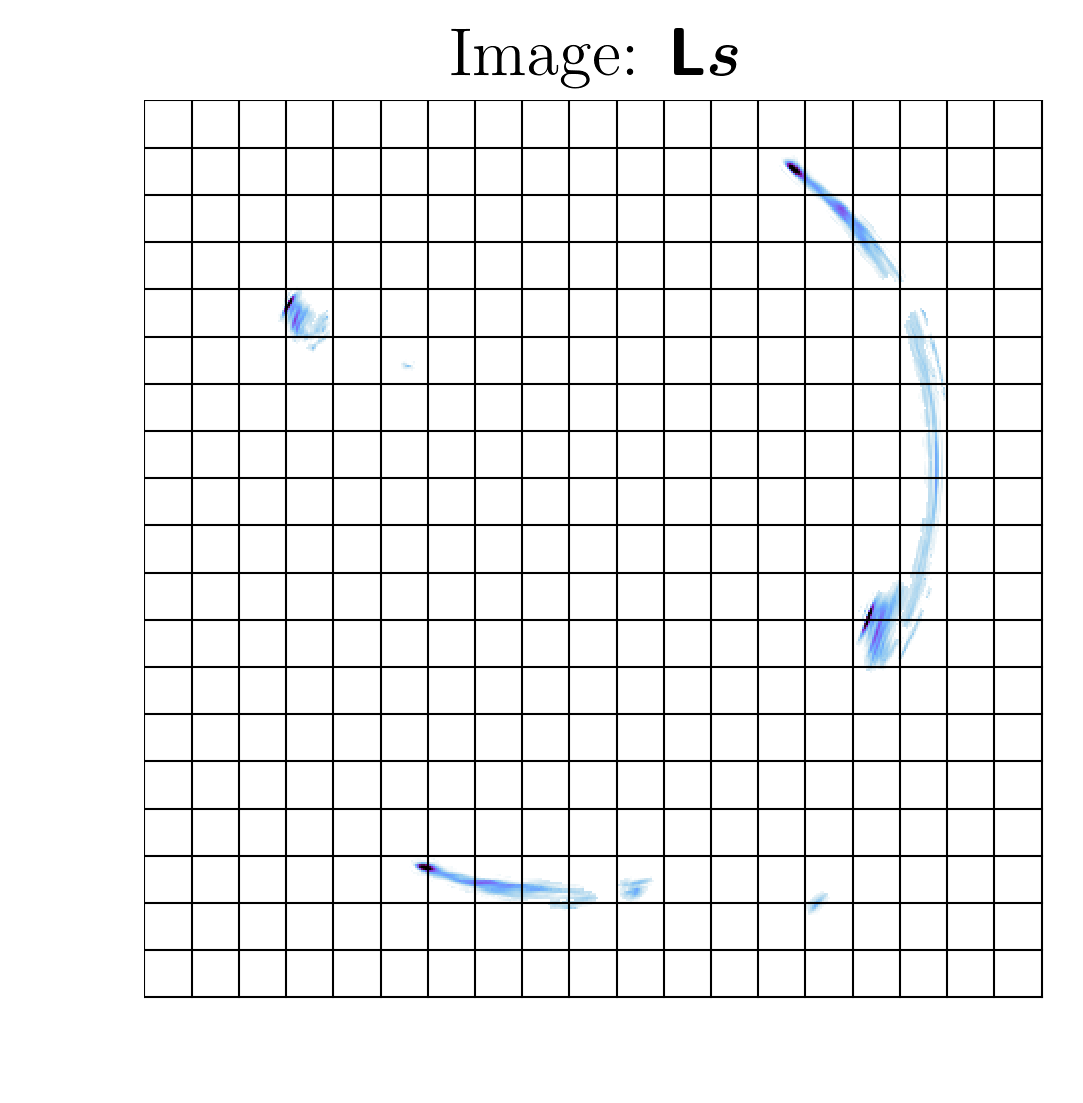}~  \includegraphics[scale=0.45]{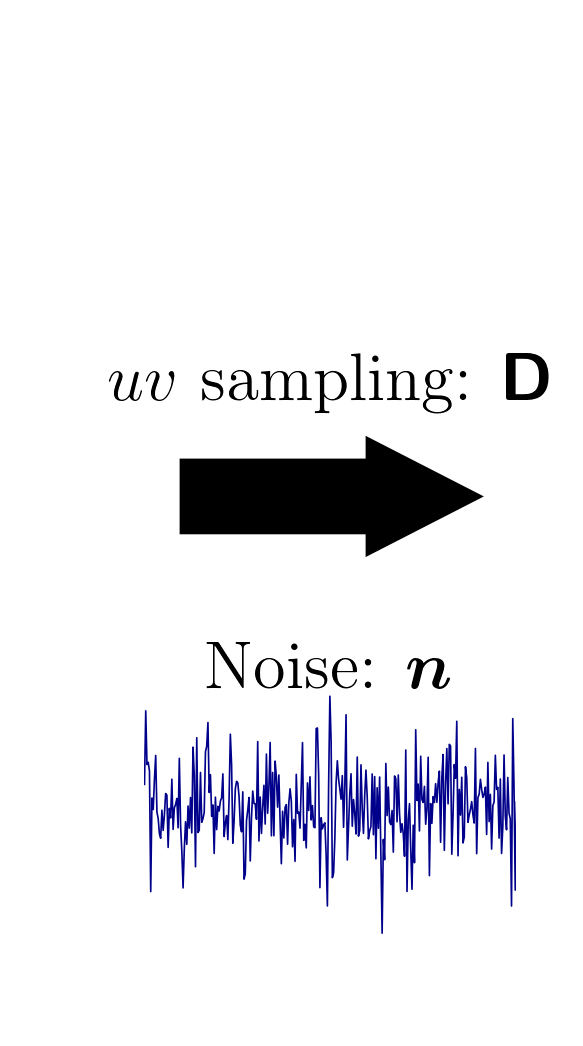}~
  \includegraphics[scale=0.45]{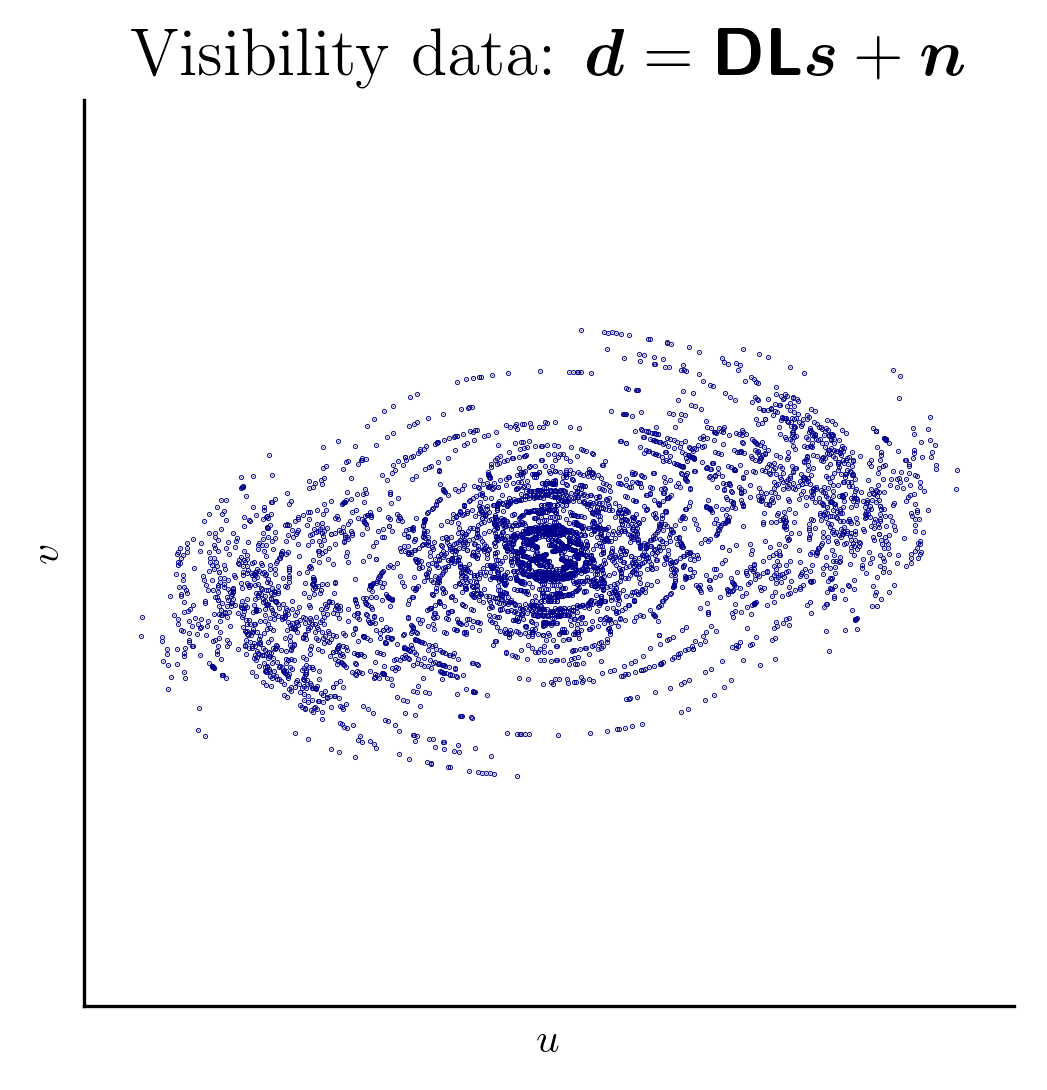}~~
  \caption{Schematic of the instrumental response described in Section \ref{sec:dataresponse}. Left panel: the source brightness $\source$ is discretized onto an adaptive Delaunay-tesselated mesh as in \protect\cite{vegetti2009}. Left-middle panel: the gravitational lens $\lensop(\etalens)$ maps emission from the source plane to the pixellated image plane according to the lens equation, which is calculated from a lens mass model that is parameterized by $\etalens$ (see Section \ref{sec:lensmodel}). Right-middle panel: this sky-plane is then sampled by the interferometer, which observes Fourier modes in the $uv$ plane, we denote this non-uniform Fourier transform as $\dft$; the data also contain additive instrumental noise $\noise$. Right panel: the full response of the data to the source brightness, gravitational lens, and interferometer is then $\data = \dft \lensop (\etalens) \source + \noise$, where the data are interferometric visibilities.}
  \label{fig:opflow}
\end{figure*}

For the analysis presented here, we use an extended version of the Bayesian modelling framework developed by \cite{koopmans2005} and \cite{vegetti2009}. This method discretizes the source plane using a Delaunay tessellation generated by rays cast backwards from the lens plane, so that the source plane resolution is naturally adapted to the lensing magnification (e.g. see \citealt{vegetti2009}). It was more recently modified to handle the case of multi-channel optical/infrared data by \cite{frizzo2018}, leveraging the magnification introduced by lensing to study the kinematic properties of sources at high redshift. 

In this section, we review the Bayesian inference framework used to find the source brightness distribution and fit lens parameters. An introduction to lens modelling in a Bayesian framework is given by \cite{suyu2006}, while a detailed discussion of the modelling code from which this work descends is described by \cite{vegetti2009} and \cite{frizzo2018}. For a general overview of hierarchical Bayesian inference, we refer the reader to \cite{mackay1991}. However, we deem it important to re-introduce the mathematical notation here in the context of radio interferometry. This will aid the explanation of the numerical methods used to accelerate these computations for large data sets in Section \ref{sec:radioimager}.  The next sections introduce many new vectors and operators, so for the reader's convenience we give a summary of our notational conventions in Table \ref{tab:notation}.

\begin{table}
\centering
    \begin{tabular}{c c p{4cm}}
    \hline
     Object & Dimension & Description \\
     \hline
      $\nvis$ & Scalar & Number of Fourier-plane visibilities \\
      $\ngrid$ & Scalar & Number of image-plane pixels \\
      $\nsource$ & Scalar & Number of source-plane vertices \\
      $\data$ & $2 \nvis$ & Data (complex visibilities) \\
      $\noise$ & $2 \nvis$ & Noise (for real and complex components of the data) \\
      $\noisecov$ & $2 \nvis \times 2 \nvis$ & Noise covariance matrix \\
      $\source$ & $\nsource$ & Source surface brightness \\
      $\etalens$ & $\mathcal{O}(10)$ & Lens parameters \\
      $\lensop(\etalens)$ & $\ngrid \times \nsource$ & Lensing operator \\
      $\dft$ & $2\nvis \times \ngrid$ & Nonuniform discrete Fourier transform (NUDFT) \\
      $\regul$ & $\nsource \times \nsource$ & Source regularization operator \newline (curvature, gradient, or brightness) \\
      $\msol$ & $\nsource \times \nsource$ & Maximum a posteriori source inversion matrix \\
      $\lams$ & Scalar & Source regularization strength \\
      $\nufft$ & $2\nvis \times \ngrid$ & Nonuniform fast Fourier transform (NUFFT) \\
      $\imcov$ & $\ngrid \times \ngrid$ & Image-plane noise covariance operator \\
      $\alpha$ & Scalar & NUFFT oversampling ratio \\
      $\dbconv$ & $2 \alpha^2 \ngrid \times 2 \alpha^2 \ngrid$ & Fourier-plane dirty beam convolution matrix \\
      $\fft$ & $2 \alpha^2 \ngrid \times \alpha^2 \ngrid$ & Fast Fourier transform (FFT) \\
      $\gridder$ & $2\nvis \times 2 \alpha^2 \ngrid$ & De-gridding operator \\
      $\zpad$ & $\alpha^2\ngrid \times \ngrid$ & Mask and zero-padding operator \\
      $\apod$ & $\ngrid \times \ngrid$ & Apodization correction operator \\
      \hline
    \end{tabular}
    \caption{Summary of the mathematical notation used in this paper. Note that for complex quantities, we introduce an extra factor of two in the dimensions. Likewise, the operators $\fft$, $\gridder$, and $\zpad$ act on a zero-padded image with an oversampling factor of $\alpha$ in each image dimension, increasing the number of image pixels to $\alpha^2 \ngrid$. We use $\alpha=2$ for this work, but keep the notation unchanged for generality.}
    \label{tab:notation}
\end{table}

\subsection{Data response}
\label{sec:dataresponse}

The data in question for this paper are radio interferometric visibilities. Each data point consists of a measurement of a complex number corresponding to a sample of the Fourier-transformed sky brightness (the $uv$ plane).  We denote the data vector hereafter as $\data$, the set of all radio visibility measurements. We assume uncorrelated Gaussian noise $\noise$ in the visibility data, so that the noise covariance is given by a diagonal matrix $\noisecov$ \citep[see e.g.][]{Wucknitz02,Thompson2017}. We denote the surface brightness in the source plane by the vector $\source$, which contains brightness values at each vertex of the Delaunay tessellation in the source plane discretization developed by \cite{vegetti2009}. 

The lens is described by the vector $\etalens$, which parameterizes the projected surface mass density of the lensing galaxy; it typically contains $\mathcal{O}(10)$ values (see Section \ref{sec:lensmodel} for details). 
$\etalens$ may also contain subhaloes and/or line-of-sight haloes.  Here we consider only a smooth lensing potential, reserving the study of these low-mass haloes for future work. The lensing operator itself is a matrix mapping the source light $\source$ to the image plane, which is a uniform Cartesian grid, with a number of pixels set by the user. It is important to note that in constructing the lensing operator $\lensop$, we do not compute deflection angles for every image-plane pixel. Rather, we mask the image plane such that all pixels outside of the mask are assumed to have zero brightness. We describe our method for creating this mask in Section \ref{sec:mask}. We emphasize that $\lensop(\etalens)$ is a function of the lens parameters $\etalens$, but for compactness we write it simply as $\lensop$ where appropriate. The sky brightness is then given by $\lensop \source$.

The operator of most consequence for working with interferometric data is $\dft$, which simulates the response of an interferometer by Fourier transforming the pixellated sky brightness distribution into a set of complex visibilities. It is important to note that the visibilities are not sampled at evenly spaced points in the $uv$ plane, meaning that formally $\dft$ is a nonuniform discrete Fourier transform (NUDFT or in brief DFT). Symbolically, $\dft$ is trivial to write down. However, it is a dense matrix operator, which presents practical challenges due to memory and speed limitations.  

When we include the noise, we obtain the response of the data to a given source brightness distribution, gravitational lens, and interferometer configuration:
\begin{equation}
    \data = \dft \lensop(\etalens) \source + \noise\,.
\end{equation}
We show a schematic of the data response in Fig.~\ref{fig:opflow}. As we describe in the next section, our goal is to jointly infer $\source$ and $\etalens$ from the observed data $\data$, but without fitting to $\noise$. This is done by maximizing the posterior probability of the model parameters given the data, as described in more detail in the following sections.
 
\subsection{Source inversion} 
\label{sec:inversion}

The first level of the inference process is the source inversion. Given a set of fixed lens parameters $\etalens$, source regularisation level $\lams$ and the data $\data$, we wish to infer the best source $\source$. In the language of Bayes' theorem, we wish to maximize the posterior probability 
\begin{equation}
    P(\source\mid\data,\etalens,\lams) = \frac{ P(\data\mid\source,\etalens) \, P(\source\mid\lams)}{ P(\data|\etalens,\lams) }\,,
\end{equation}
where $P(\data\mid\source,\etalens)$ is the likelihood and $P(\source\mid\lams)$ is a prior distribution on $\source$. The denominator $P(\data|\etalens,\lams)$ is independent of  $\source$ so we may ignore it when maximizing the posterior. It is a normalization factor known as the \emph{evidence} that takes on a special role in hierarchical Bayesian inference, as it allows to statistically compare different models in a way that automatically takes into account Occam's razor.

We assume a Gaussian likelihood formed by considering the forward-modelled visibilities $\dft\lensop\source$ obtained from the model source $\source$. The log-likelihood is then
\begin{equation}
    P(\data \mid \source,\etalens) = \frac{1}{Z_D} \exp \left (-\frac{\chi^2}{2}\right)\,,
\end{equation}
where
\begin{equation} \label{eq:chi2}
    \chi^2 = (\dft\lensop\source-\data)^T\noisecov(\dft\lensop\source-\data)
\end{equation}
and $Z_D = \sqrt{\det(2\pi \oper{C})}$ is the normalization.

A fundamental feature of radio interferometry is that the $uv$ plane is sampled only at a finite number of points. Therefore, in order to optimize our model for $\source$ and $\etalens$ we must introduce some extra information in the form of a prior. We choose the source prior such that either the total curvature, gradient, or magnitude of the resulting surface brightness distribution is minimized. The form of the prior is fixed to one of these three choices, which we denote by the discrete operator $\regul$. This is fixed throughout the entire inference process, though the Bayesian evidence for different choices of $\regul$ can be used to set the best form of regularization given the data. The strength of the regularization is given by the scalar $\lams$. We choose a Gaussian form for the prior as well, which is
\begin{equation}
 P(\source \mid \lams) = \frac{1}{Z_R} \exp \left [ -\frac{1}{2}\lams(\regul \source)^T \regul \source \right]\,,
\end{equation}
with normalization $Z_R = \sqrt{\det[2\pi(\lams \regul^T \regul)^{-1}]}$.

We now solve for $\source$ such that we maximize the posterior probability for a set of fixed lens parameters $\eta$ and regularization constant $\lams$. It is more convenient to work in terms of the log-posterior
\begin{multline}
   \log P(\source \mid \data, \etalens, \lams) = -\frac{1}{2}\left[(\dft\lensop\source-\data)^T\noisecov(\dft\lensop\source-\data) + \lams(\regul \source)^T \regul \source \right]\\ 
   -\frac{1}{2}\logdet(2\pi \oper{C}) -\frac{1}{2}\logdet[2\pi(\lams \regul^T \regul)^{-1}]\,.
\end{multline}
Setting $\frac{\partial}{\partial \source} \log P(\source\mid\data,\etalens,\lams) = 0$ yields the regularized least-squares equation
\begin{equation} \label{eqn:leastsquares}
   \left[(\dft\lensop)^T\noisecov \dft\lensop + \lams \regul^T \regul \right] \smp = (\dft\lensop)^T \noisecov \data\,.
\end{equation}
For convenience of notation throughout this paper, we denote the maximum a posteriori source inversion matrix or briefly the solution matrix for $\smp$ as
\begin{equation}
    \msol \equiv \left[(\dft\lensop)^T\noisecov \dft\lensop + \lams \regul^T \regul \right]\,.
\end{equation}
Equation \eqref{eqn:leastsquares} gives the maximum a posteriori (MAP) source $\smp$; it is simply a linear system with dimensions $\nsource \times \nsource$. However, a major challenge arises when considering that the operator $\dft$ from which $\msol$ is composed is dense and contains $2\nvis \times \ngrid$ elements. This precludes the use of a direct solver when either $\nvis$ or $\ngrid$ is large. We dedicate Section \ref{sec:radioimager} to describing the numerical methods used to solve this system in practice.

\subsection{Lens parameters and regularization} 
\label{sec:lensparams}

In the second level of inference, we optimize the lens model $\etalens$ and the regularization strength $\lams$. This is achieved by maximizing
\begin{equation}\label{eq:bayev}
P(\etalens,\lams\mid\data) = \frac{ P(\data\mid
		\etalens,\lams) \, P(\etalens) \, P(\lams)}{ P(\data)}\,.
\end{equation}
The priors $P(\etalens)$ and $P(\lams)$ are chosen by the user; in this work we use a uniform prior on $\etalens$ and a log-uniform prior on $\lams$. The normalization $P(\data)$ is constant in this step, so we may ignore it for now. The quantity $\log{P(\data \mid \etalens, \lams)}$ is given by
\begin{multline} \label{eq:baypos}
    	2\log{P(\data \mid \etalens, \lams)} = -\chi^2 - \lams \smp^T \regul^T \regul \, \smp - \logdet \msol \\
    	+ \logdet (\lams \regul^T \regul)  + \logdet (2\pi \noisecov)\,,
\end{multline}
where the $\chi^2$ takes the form of equation \eqref{eq:chi2} using the MAP source $\smp$ computed at every step using equation \eqref{eqn:leastsquares}. For a derivation of this expression, see \cite{suyu2006}, and \cite{frizzo2018} for an extension to the three-dimensional domain.

We maximize equation \eqref{eq:bayev} using a simulated annealing routine. This algorithm does not require gradients to operate, which are difficult to compute for the nonlinear parameters $\etalens$. In practice, we do not optimize for both $\etalens$ and $\lams$ simultaneously. Instead, we first set $\lams$ to an artificially large value in order to obtain an over-regularized source inversion. This smooths the posterior landscape and allows the lens parameters $\etalens$ to closely approach their best values while initially avoiding local minima. We then alternate optimizing for $\etalens$ and $\lams$ until convergence is reached for both. The choice to not simultaneously optimize $\etalens$ and $\lams$ is good practice for Bayesian modelling techniques in general; see for example, \cite{mackay1991}. At each step of the optimisation, the corresponding most probable source is obtained by solving the linear system introduced in the previous section.

Of special importance in the expression above is the term $\logdet \msol$, which normalizes the posterior from the previous inference step. As we discuss in Section \ref{sec:radioimager}, we cannot build $\msol$ explicitly for large problem sizes. We instead must find a sufficiently accurate and computationally tractable approximation to this log-determinant term, which we present in Section \ref{sec:logdet}.

\section{NUFFT-based radio imaging} 
\label{sec:radioimager}

Equation (\ref{eqn:leastsquares}) is mathematically straightforward. However, when the data dimension $\nvis$ becomes large, the solution of this linear system becomes intractable for a direct linear solver. This is because the direct Discrete Fourier Transform (DFT) operator $\dft$ contains $2 \nvis \times \ngrid$ nonzero elements. A conservative $\ngrid = 128^2$ pixels and $\nvis = 10^7$ visibilities at single precision would demand 1.2 TB simply to store in memory. Even if memory were not a constraint (and indeed the matrix elements could be evaluated on the fly) a matrix-vector multiplication by $\dft$ would cost $3.3\times 10^{11}$ floating point sine and cosine evaluations.

A further point is that for high-resolution VLBI observations, a large number of pixels is needed to correctly sample the model image and avoid aliasing artifacts. $\ngrid$ is a somewhat free parameter: it is set by the user, but the choice of $\ngrid$ must be informed by the Nyquist frequency of the grid and the $uv$ coverage of the observation (see Section \ref{sec:mask}). Assuming that $\msol$ can be assembled, a dense matrix with dimensions $\nsource \times \nsource$ rapidly becomes computationally intractable, especially when evaluating many solutions during Monte Carlo sampling.

We now introduce our novel approach for modelling large interferometric data sets. We first replace the direct DFT operator $\dft$ with a non-uniform Fast Fourier Transform (FFT) operation $\nufft$ (NUFFT hereafter), which we discuss in Section \ref{sec:nufft}.
We also replace the image-plane noise covariance $\imcovexplicit = \dft^T \noisecov \dft$ with an FFT-based convolution operation $\imcov$, which we introduce in Section \ref{sec:imcov}. The replacement of true matrices with operators (which cannot explicitly be inverted) requires the use of an iterative solver, which we discuss in Section \ref{sec:cgsolver}. 

\subsection{The NUFFT operator} 
\label{sec:nufft}

The NUFFT operator replaces the dense NUDFT matrix $\dft$ with a composition of functions that are fast to evaluate.  \revisions{In the order that they are applied rightward to an image, these are an apodization correction $\apod$, a zero-padding/masking operation $\zpad$, an FFT $\fft$, and a de-gridding operation $\gridder$.} Thus, our NUFFT operator takes the form
\begin{equation}
    \nufft = \gridder \fft \zpad \apod\,.
\end{equation}
The FFT $\fft$ is memory-efficient and runs in $\mathcal{O}(\ngrid \log \ngrid)$ time. Since $\fft$ operates on a regular Cartesian grid, we must interpolate the visibilities off of a pixellated $uv$ plane using the de-gridding operator $\gridder$. The apodization correction $\apod$ is the inverse Fourier transform of the gridding kernel. \revisions{It compensates for the effect of the gridding kernel's shape in the final interpolated visibilities}. Finally, the zero-padding operation $\zpad$ masks aliasing errors around the edges of the image plane; this is equivalent to oversampling the Fourier transform of the image by a factor $\alpha$. We use a pad factor (or, equivalently, oversampling ratio) of $\alpha=2$. This is a common choice in NUFFT implementations, but it is also necessary for performing non-periodic convolutions with the dirty beam in the image plane (see Section \ref{sec:imcov}).

The operator $\gridder$ is responsible for interpolating the arbitrarily-spaced visibilities onto or off of regular sampling locations in the $uv$ plane. For example, consider a visibility $\data_i$ (a complex number) that lies at the arbitrary $uv$ coordinate $\uu_i$. We impose a regular Cartesian grid onto the $uv$ plane using $\kk_j$, with points spaced at regular intervals $\Delta k$. Then the gridding operation becomes
\begin{equation}
    (\gridder^T \data)_i = \sum_j K\left(\frac{\uu_j-\kk_i}{\Delta k}\right) \data_j\,,
\end{equation}
where $K(\cdots)$ is the gridding kernel, with a finite support radius, meaning that it is compact in the $uv$ plane. We use the Kaiser-Bessel kernel \citep{kaiser1980} with a support radius $\wsup=4$ grid-points. This kernel has been shown to provide near-optimal image reconstruction, in the sense that it minimizes aliasing errors. It also uses only analytic functions, which simplifies the implementation. \cite{beatty2005} provide both an excellent explanation of the NUFFT in general, as well as best practices for using this kernel. We discuss this kernel and its Fourier transform (the apodization correction $\apod$) further in Appendix \ref{app:kb}. We note that the choice of zero-padding factor $\alpha$ and kernel support radius $\wsup$ can be chosen to provide an arbitrary level of accuracy; we find that the values used for this paper work well for our purposes.

We accelerate the operator $\gridder$ using a GPU compute kernel written in the \software{cuda} language. The gridding operation is nearly embarrassingly parallel, so the \software{cuda} kernel essentially consists of a simple loop unrolling with no further special treatment needed.

Our intent here is simply to introduce the operators involved in the NUFFT operation $\nufft$, the most important of which are the FFT $\fft$ and the de-gridding operator $\gridder$. For more detailed information, the reader is directed towards the abundant literature on the theory of the NUFFT \citep[e.g.][]{jackson1991, fessler2003, greengard2004, beatty2005}.

\subsection{Image-plane noise covariance} 
\label{sec:imcov}

We group together the operators $\dft^T \noisecov \dft \equiv \imcovexplicit$ from equation \eqref{eqn:leastsquares} to form the noise covariance in the image-plane basis. Despite being fully dense in its explicit matrix representation, $\imcovexplicit$ can be applied efficiently using a FFT. It is straightforward to show that for any two image-plane pixels at locations $\xx_i$ and $\xx_j$, the noise covariance between the two is
\begin{equation} \label{EQ:DBCONV}
    \oper{C}_{x,ij}^{-1} = \sum_{k} \, \frac{1}{\sigma_k^2} \cos\left[2\pi \uu_k \cdot (\xx_i-\xx_j) \right]\,,
\end{equation}
where the sum is over all visibilities: here, $\uu_k$ is the $uv$ coordinate of the $k^\mathrm{th}$ visibility, and $\sigma_k$ is the noise covariance of its real part. In other words, $\imcovexplicit$ performs a convolution with the naturally-weighted dirty beam. We therefore replace $\imcovexplicit$ with the operator $\imcov$, which duplicates the action of a matrix-vector multiplication by $\imcovexplicit$. To apply $\imcov$ in practice, we must only compute the naturally-weighted dirty beam and perform a convolution using an FFT. For a short derivation of equation \eqref{EQ:DBCONV}, see Appendix \ref{app:dbder}.

We compose $\imcov$ as follows. Upon first initializing the modelling code, we compute the dirty beam using the same pixel scale, but twice the physical extent of the image-plane grid, so that $N_\mathrm{grid,dirty~beam} = \alpha^2 \ngrid$. We do this using the NUFFT as described in Section \ref{sec:nufft}, simply doubling the number of pixels in each direction. We then apply the FFT $\fft$ to the dirty beam and store it in the diagonal matrix $\dbconv$.
To apply $\imcov$, the image is first zero-padded using $\zpad$. As the image plane is non-periodic, our use of the zero-padding factor $\alpha=2$ comes into play, as it prevents spurious periodic correlations from being introduced in the image plane. We then FFT the zero-padded image, apply the Fourier-transformed dirty beam $\dbconv$, apply the inverse FFT, and remove the zero-padding. In our operator notation, this is
\begin{equation} \label{eq:cxinvop}
    \imcov \equiv \zpad^T \fft^T \dbconv \fft \zpad\,.
\end{equation}

\subsection{Iterative solution and preconditioning}
\label{sec:cgsolver}

Making the substitutions $\dft \rightarrow \nufft$ and $\imcovexplicit \rightarrow \imcov$ gives the NUFFT-based version of equation \eqref{eqn:leastsquares},
\begin{equation} \label{eq:nufftlsq}
   \left[\lensop^T \imcov \lensop + \lams \regul^T \regul \right] \source = \lensop^T \nufft^T \noisecov \data\,,
\end{equation}
which we will refer to later in this section. The consequence of introducing the operators $\nufft$ and $\imcov$ is that we no longer have an explicit matrix representation for this equation. Rather, the left-hand side exists only as a function that emulates matrix-vector multiplications.  For the rest of this section, we describe in more detail our handling of the linear solver, and the computation of the log-determinant (see Section \ref{sec:logdet}) under this restriction.

The only way of obtaining a solution is to use an iterative linear solver. In general terms, such solvers apply a linear operator repeatedly, subtracting residuals from a trial solution until a desired tolerance is reached. We have adopted a ubiquitous choice, the preconditioned conjugate gradient solver (PCCG), with a convergence tolerance of \revisions{$10^{-8}$ (a smaller value would be unnecessary since we are fundamentally limited in accuracy by the NUFFT operator; see Section \ref{sec:T1})}. The conjugate gradient method is derived in such a way that the largest residual components are subtracted first, giving fast convergence. The use of a preconditioner further accelerates convergence.  We use the PCCG solver implementation provided by the \software{petsc} framework \citep{petsc-efficient, petsc-user-ref}, which is an MPI-parallel library designed for solving large linear systems. For a general linear system $\msol \source = \bb$, the preconditioner is a matrix \emph{approximating} $\msol$ so that we can instead solve
\begin{equation}
	\pc^{-1} \msol \source = \pc^{-1} \bb\,.
\end{equation}
If $\pc^{-1} \msol \approx \oper{I}$, then a solution can be achieved in far fewer iterations. Our source inversion absolutely depends on finding a good preconditioner matrix, as the original system, given in equation \eqref{eqn:leastsquares}, can have condition numbers of higher than $10^{10}$ depending on the $uv$ coverage and regularization strength (\emph{condition number} is a measure of how singular a matrix is; the identity matrix $\oper{I}$ has a condition number of 1).

Finding a suitable preconditioner matrix $\pc$ is highly dependent on the features of the particular problem under consideration. Ours is based on the image-plane noise covariance $\imcovexplicit = \dft^T \noisecov \dft$. As discussed in Section \ref{sec:imcov}, each row of this matrix simply contains the naturally-weighted dirty beam; that is, it is dominated by its diagonal when the $uv$ coverage is good.

We take advantage of this property by approximating this matrix with its diagonal, such that each entry is simply the brightest pixel in the dirty beam. Using only the diagonal of $\imcovexplicit$ (rather than e.g. three elements per row) guarantees positive-definiteness of $\pc$, which is a requirement for CG solvers. Our preconditioner is then
\begin{equation} \label{eq:pc}
	\pc = \left(\sum_{k} \, \frac{1}{\sigma_k^2} \right) \lensop^T \lensop + \lams \regul^T \regul\,,
\end{equation}
which is a sparse matrix of dimension $\nsource\times\nsource$ that can be computed explicitly. We apply a Cholesky decomposition to $\pc$, which we compute once for each source inversion. We use the \software{mumps} direct solver \citep{MUMPS01,MUMPS02} for this decomposition, which is conveniently provided within \software{petsc}. This decomposition is equivalent to $\pc^{-1}$ and can be quickly applied at each CG iteration. We find that this preconditioner works extremely well, reducing the condition number by a factor of $\sim 10^6$ or more and requiring $\sim 100\times$ fewer iterations to achieve convergence for the problems presented in this paper. \markup{We note that the choice of preconditioner determines the speed with which convergence is reached and not the precision of the solution.}

\subsection{Log-determinant approximation} 
\label{sec:logdet}

As we point out in Section \ref{sec:lensparams}, we also must compute the log-determinant of the matrix $\msol$. This is again problematic, as we do not possess this matrix explicitly. We note that there exist methods to estimate log-determinants of matrices based on power series expansions \citep[e.g.][]{han2015,fitz2017,granz2018}. However, these methods rely on stochastic trace estimations using random probing vectors. The stochastic nature of such determinant estimators poses a problem for Monte Carlo samplers such as \software{multinest}~\citep{feroz2009}, which require a deterministic and reasonably smooth posterior landscape to explore. 

However, we obtain $\log \det \pc$ at no extra computational cost when we compute the Cholesky decomposition of $\pc$ in preparing to apply the preconditioner. We have found that this provides an approximation to the log-determinant that is sufficiently accurate when we test this on simulated observations (see Section \ref{sec:ldtests}).

\section{Generating Simulated Observations} 
\label{sec:modelling}

 \begin{figure*}
 \centering
   \includegraphics[scale=0.6]{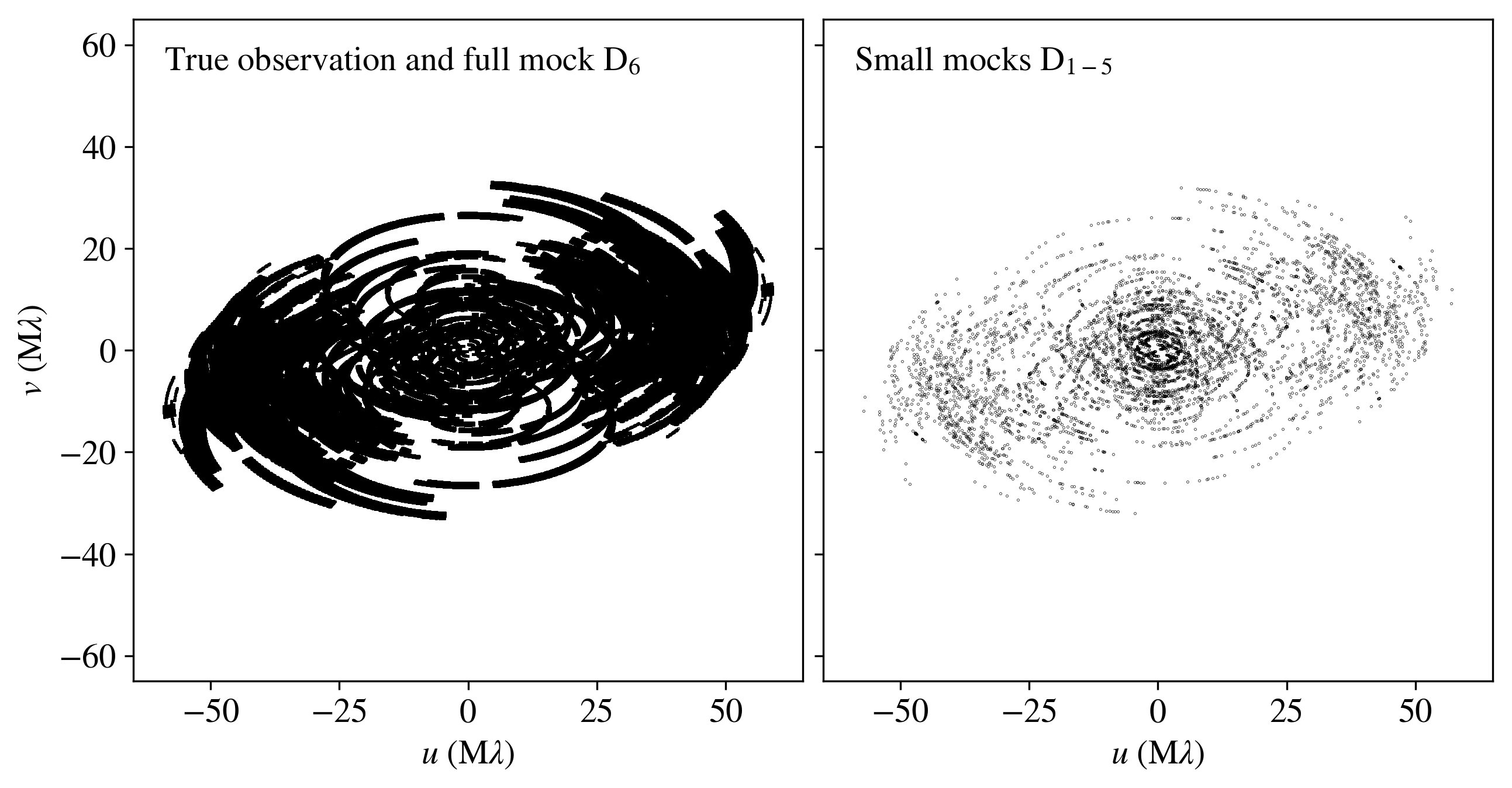}
  \caption{Left panel: the $uv$ coverage of the global VLBI observation of the gravitational lens system MG J0751+2716 at 1.65~GHz, described in Section \ref{sec:modelling} and for the mock data set $\rm D_6$ introduced in Section \ref{sec:simobs}. Right panel: the $uv$ coverage of the small mock data sets $\rm D_1$ to $\rm D_5$ presented in Section~\ref{sec:simobs}.}
  \label{fig:uvcoverage}
\end{figure*}

Our goal in developing the novel methodology described above, is to model high angular-resolution VLBI observations of gravitationally lensed sources that until now had a prohibitively large number of visibilities for modelling using standard DFT approaches. To this end, we derive a mock data set from actual global VLBI observations of the gravitational lens system MG J0751+2716, which we use in Section \ref{sec:simobs} to test the robustness of our modelling approach. In this section, we describe the data preparation, noise estimation, and other preliminary steps for verifying our modelling technique on simulated data.

\subsection{Data preparation and flagging}

MG J0751+2716 is a gravitationally lensed radio-loud quasar at redshift $z_\mathrm{src}=3.2$ \citep{tonry1998,alloin2007}. The lensing galaxy lies at $z_\mathrm{lens}=0.35$ \citep{tonry1998, momcheva2006}, and is part of a group of galaxies that introduces a strong external shear to the lens model \citep{lehar1997,momcheva2006, alloin2007}. The properties of this lens system are well-studied; the most recent investigation has been carried out by \cite{spingola2018}, to which we refer for further details. 

The global VLBI observation of MG J0751+2716 used in this paper was undertaken on 2012 October 12 for a total of 18.5 hours (PI: McKean; ID: GM070). The observation was performed using 13 antennas from the European VLBI Network (EVN), 10 antennas from the Very Long Baseline Array (VLBA) and the Green Bank Telescope (note that not all antennas were used simultaneously for the whole observation due to hour-angle restrictions). The data were taken in dual-polarization mode (RR and LL correlations) at a central observing frequency of 1.65 GHz with a total bandwidth of 64 MHz, which was divided into 256 spectral channels. The visibilty averaging time was 2~s. The data were processed by \cite{spingola2018}, whereupon most of the radio frequency interference (RFI) was removed (flagged) from the data set. The visibilities were calibrated following standard VLBI procedures for phase referenced observations, and a detailed explanation of the methods used is presented by \cite{spingola2018}. 

\revisions{We inspected the calibrated visibilities in order to further identify and remove any remaining RFI from the data.} In addition to that identified by \cite{spingola2018}, we also found other RFI between 1649 and 1652 MHz for Jodrell Bank, between 1676 and 1678 MHz and 1686 and 1688 MHz for Medicina, and between 1658 and 1661 MHz for Noto. We note that the latter antenna also had an unusually higher noise temperature relative to the rest of the array.  We also found numerous other regions in time and frequency where transient RFI were present. Instead of combing through and manually flagging problematic data, we impose a cut in which all visibilities with noise exceeding 1 Jy are flagged. We find that this removes the aforementioned RFI and, after these flagging steps, the observation contains $2.4 \times 10^8$ visibilities ($\sim$70 per cent of the initial visibilities).

\subsection{Noise estimation} 
\label{sec:noiseest}

We compute the noise directly from the visibility data as follows. First, we split the data by baseline, channel, and correlation.  We further divide each of these data blocks into 20-minute time intervals. Within each of these intervals, we subtract time-adjacent visibilities from one another in order to remove the sky signal, under the reasonable assumption that the visibility function is slowly varying in time. We then compute the RMS of the result, \revisions{scaled by $\sqrt{2}$ to account for the subtraction}, and assign it as the noise for that interval. We insert these noise values back into the original measurement set for use in the lens modelling process.

After estimating the noise from the raw data, we save the tables to disk for further use in our code verification tests. This proves useful because it gives us the ability to draw random noise realizations that are consistent with the original observation, and apply them to simulated observations (see Section \ref{sec:simobs}). 

\subsection{Image-plane grid} 
\label{sec:mask}

Although the data dimensions $\nvis$ are fixed by the observation, we must determine appropriate dimensions for the image-plane pixellization $\ngrid$ and the source-plane discretization $\nsource$. We set $\ngrid$ based on the $uv$ coverage of the data and the Nyquist criterion. The maximum $uv$ distance in the observation is $59~\mathrm{M}\lambda$ in approximately the east--west direction (see Fig.~\ref{fig:uvcoverage}), corresponding to an angular resolution of 3.5 mas. The lens system has an Einstein radius of $\sim 0.5$ arcsec, so we choose $\ngrid = 1024^2$ with a spatial extent of $1.5\times1.5~\mathrm{arcsec^2}$ that is centred on the mid-point between the lensed images. This gives a pixel size of 1.5 mas, which places the Nyquist frequency safely above the angular resolution of the observation and avoids aliasing in the image. 

We create the image-plane mask by starting from the \software{clean}ed image of MG J0751+2716 that was made by \cite{spingola2018}. We load the image, and apply a threshold of $5\sigma_\mathrm{RMS}$, where $\sigma_\mathrm{RMS}\simeq41 \mu$Jy~ beam$^{-1}$ is the RMS residual off-source noise. We then pad around these pixels by three beam-widths $3 \times 5.7$ mas$^2$ in all directions.  This procedure yields a tight mask (see Section \ref{sec:fullmock}) while ensuring that all statistically-significant emission is included in the lens modelling process. The source-plane discretization is determined by the number of unmasked pixels from which the lensing operator $\lensop$ is computed; for this mask there are $\nsource=31959$ vertices.

\begin{table} 
\centering
\begin{tabular}{c l l }
\hline
Name & $\nvis$ & SNR\\
\hline
$\rm{D_1}$& $2.8\times10^3$ &$4.49\times10^3(\equiv\snrfid)$\\
$\rm{D_2}$& $2.8\times10^3$ &$\snrfid\times 10$\\
$\rm{D_3}$& $2.8\times10^3$ &$\snrfid\times 10^{-1}$\\
$\rm{D_4}$& $2.8\times10^3$ &$\snrfid\times 10^{-2}$\\
$\rm{D_5}$& $2.8\times10^3$ &$\snrfid\times 10^{-3}$\\
$\rm{D_6}$& $2.4\times 10^8$ & $4.43\times10^3$ \\
\hline 
\end{tabular} 
\caption{Summary of all the simulated observations used to test our method in Section \ref{sec:simobs}. For each we provide the number of visibilities and the signal-to-noise ratio.}
\label{tab:mocks}
\end{table}

\begin{table*} 
\centering
\begin{tabular}{c l l l l l}
\hline
Test& Data Set& Focus & $\eta$ & $\lams$ & Section\\
\hline
T1 & $\rm{D_1}$ & NUFFT operator & N/A & N/A & \ref{sec:T1}\\
T2 & $\rm{D_1}$ & Preconditioner and iterative solution& $\equiv\etagt$&$\equiv \lams^{\rm DFT,MAP}$&\ref{sec:T2}\\
T3 & $\rm{D_1}$ to $\rm{D_5}$ & Log-determinant &$\equiv \etagt$ & free & \ref{sec:ldtests}\\
T4 & $\rm{D_1}$ to $\rm{D_5}$ & Full modelling & free & free &\ref{sec:snr} \\
T5 & $\rm{D_6}$ & Image-plane mask & $\equiv\etagt$ & free & \ref{sec:masktest}\\
T6 & $\rm{D_6}$ & Full modelling & free & free & \ref{sec:fullmock}\\
\hline 
\end{tabular} 
\caption{Summary of all the tests used to verify our method. From right to left, we list the test name, the data used, which aspect of the modelling procedure is being tested, treatment of the non-linear parameters, and the section where the test is discussed.}
\label{tab:tests}
\end{table*}

\subsection{Lens modelling} 
\label{sec:lensmodel}

We parameterize the lens mass distribution using a cored elliptical power-law model plus external shear, with the normalized projected mass density given by
\begin{equation}
    \kappa(x,y) = \frac{\kappa_0\left(2-\frac{\gamma}{2}\right)q^{\gamma-\frac{3}{2}}}{2\left[ q^2 \left(x^2 + r_c^2 \right) + y^2  \right]^{\frac{\gamma-1}{2}}}\,,
\end{equation}
where $\kappa_0$ is the mass normalization, $q$ is the elliptical axis ratio, $\gamma$ is the power-law slope (with $\gamma=2$ corresponding to an isothermal power-law), and $r_c$ is the core radius.  We fix $r_c = 10^{-4}$ arcsec. We compute the corresponding deflection angles using the \software{fastell} library \citep{barkana1999}.

This mass model is translated and rotated to the lens position $(x_0,y_0)$ and position angle $\theta$. Finally, we include an external shear component defined by its strength $\Gamma_\mathrm{sh}$ and direction $\theta_\mathrm{sh}$. The free parameters in our lens model are then $\etalens = (\kappa_0,\gamma,q,\theta,x_0,y_0,\Gamma_\mathrm{sh},\theta_\mathrm{sh})$. 

\subsection{Mock observations based on a realistic source model}

In order to generate a set of realistic mock observations, we require a representative model for the background source surface brightness distribution and for the mass distribution of the lensing galaxy. We begin with the best-fit source reconstruction and lens model that is obtained by applying our modelling approach to a heavily averaged version of the global VLBI observation of MG~J0751+2716. We then mask out everything, except for the brightest emission components of this reconstructed source. The structure of the mock source is not critical; we simply wish to start with something that resembles the true observed source in brightness and morphology. However, this also results in a mock source that has a good balance between compact high brightness emission (from a radio core or hot-spot) and extended low surface brightness emission (from a radio-jet). We refer to this source and the lens parameters as the ``ground truth'' (GT) and denote them as $\sgt$ and $\etagt$.  Next, we lens the ground truth source forward using $\lensop(\etagt)$ to create a sky model.  The ground-truth source and sky model can be seen in the top row of Fig.~\ref{fig:full_inversions}.  The mock interferometric data are then produced using the Common Astronomy Software Applications package (\software{casa}; \citealt{casa2007}); we use \software{casa}'s \texttt{simulator} tool to Fourier-transform the sky model into a set of visibilities. Finally, we corrupt the simulated visibilities with different levels of Gaussian noise. From this initial simulated visibility data set, we create various sub-data sets to test various aspects of our NUFFT modelling approach, which we describe in the next section.

\section{Tests on simulated observations} 
\label{sec:simobs}

 \begin{figure*}
 \centering
  \includegraphics[scale=0.55]{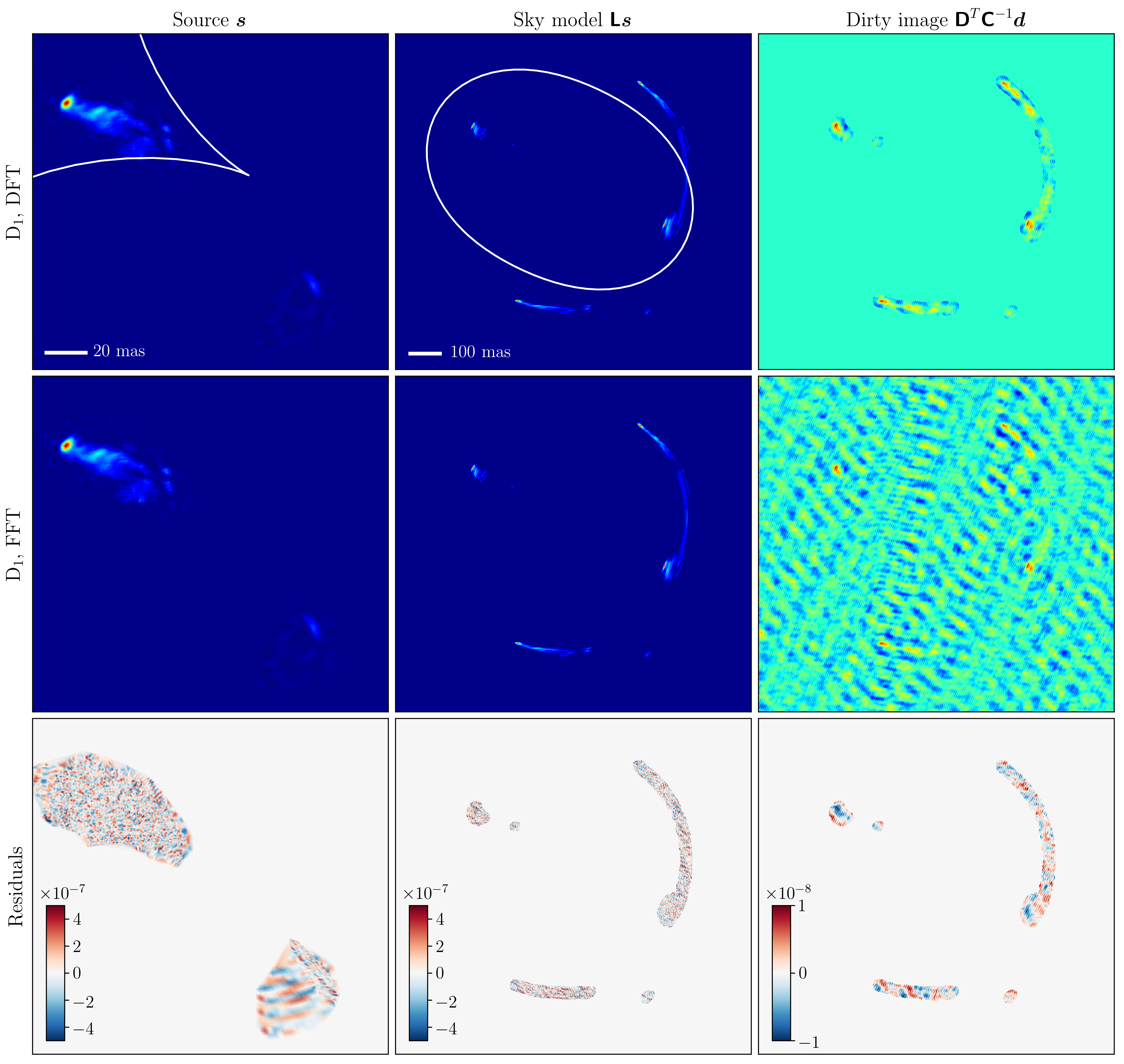}
  \caption{The results of tests T1 and T2. From top to bottom, the rows show the output from the DFT and NUFFT approach, and the difference between the two. The left (middle) column shows the maximum a posteriori source (lensed) surface brightness distribution normalized to the peak source surface brightness. The maximum error between the DFT and NUFFT  method is $7.6\times10^{-7}$, with an RMS of $1.6\times10^{-7}$. In the right column, we plot the dirty image of the data. Note that in the DFT implementation of the code, only the image-plane pixels that lie inside the mask are included as rows in the Fourier transform matrix. The maximum and RMS relative errors between the two dirty images are $1.2\times 10^{-8}$ and $3.4\times 10^{-9}$, respectively. \revisions{We also show the critical curves in the top row to illustrate the lensing configuration for the mock data (plotted in white).}}
\label{fig:dft_fft_inversions}
\end{figure*}

In this section, we test the reliability and robustness of our modelling approach by applying it to a set of simulated observations of MG J0751+2716 (see Section \ref{sec:modelling}) under different data quality assumptions. The advantage of using mock observations is that it allows for a fair comparison between the various tests and the ground-truth source and lens model.

We perform the first set of tests on mock data containing a highly reduced number of visibilities, while still matching the $uv$ coverage of the original observation as closely as possible. \revisions{For these tests, we also only cast rays from every other image-plane pixel to the source plane, interpolating as described in \cite{vegetti2009}, so that $\nsource=10464$}.  These smaller problem sizes allow us to test our matrix-free FFT implementation against a DFT-based direct solver. We generate these reduced data sets by selecting every 400\textsuperscript{th} row from a single channel of the original measurement set, leaving only $\sim 3000$ visibilities. This ensures that, even though the number of $uv$ samples is greatly reduced, the overall pattern of the visibiity sampling function is unchanged, as can be seen in Fig.~\ref{fig:uvcoverage}. 

We create five such small mock observations, which we label $\rm D_1$ to $\rm D_5$.  We corrupt $\rm D_1$ with noise that is re-scaled such that its SNR matches that of the original observation. We compute this factor via $\sqrt{N_\mathrm{vis,D_1}/N_\mathrm{vis,data}} = 275$.
We define the SNR from the ground truth source, the lens parameters, and an appropriately-scaled value of the visibility noise as
\begin{equation}
    \mathrm{SNR} \equiv \sqrt{\sum_k \frac{|\dft\lensop(\etagt)\sgt|_k^2}{\sigma_k^2} }\,,
\end{equation}
where $\sigma_k$ is the noise associated with the real or complex part of visibility $k$.
For the fiducial small mock data set, $\mathrm{SNR_{D_1}}=\snrfid\equiv4.49\times10^3$. The data sets $\rm D_2$ to $\rm D_5$ instead have SNRs that vary in factors of 10 away from $\snrfid$ in both directions.

We also generate a full mock data set from the same $uv$ coverage as the real observation itself, which we label $\rm D_6$. We add Gaussian noise as estimated from the data (see Section \ref{sec:noiseest}).  Its signal-to-noise ratio is $\mathrm{SNR_{D_6}}=4.43\times10^3$, and contains the full $2.4 \times 10^8$ unflagged visibilities from the true observation. We use this mock data set to check the ability of the modelling code to fit a source model and a set of lens parameters accurately with respect to the ground truth under the same conditions as the true observation. \revisions{We also use it to test for any potential overfitting}. A summary of the mock data sets used for these tests is presented in Table~\ref{tab:mocks}.

As described in Section \ref{sec:radioimager}, there are three substitutions that must be made in order to model large radio interferometric data sets. The first two involve the replacement of the dense DFT matrix with a NUFFT operator ($\dft\rightarrow\nufft$), and the use of a convolution with the dirty beam in place of a dense matrix multiplication ($\imcovexplicit \rightarrow \imcov$). The last is the substitution of the preconditioner log-determinant for the log-determinant of the full solution matrix. Below, we evaluate the performance of these new ingredients with a series of tests whose specifics are summarized in Table~\ref{tab:tests}.

\subsection{T1: Dirty image}
\label{sec:T1}

In this test, we verify that the operator $\nufft$ produces the same result as the DFT-based modelling code. As we want to compare directly with the DFT, we use the data set $\rm{D_1}$, and  check that the NUFFT operator computes the correct dirty image from the data.  \revisions{This is also a check that our choices of zero-padding factor $\alpha=2$ and kernel support radius $\wsup=4$ in the gridding operation provide sufficiently accurate results in the NUFFT (see Section \ref{sec:nufft} and Appendix \ref{app:kb}).}

In our notation from Section \ref{sec:radioimager}, this amounts to computing $\nufft \noisecov \data$ in the NUFFT case, and $\dft \noisecov \data$ for the DFT matrix. The dirty image of the data is an important ingredient of our inference scheme, as it is directly related to the computation of the most probable source via the right-hand side of equation (\ref{eq:nufftlsq}). The results of this test are contained in the rightmost column of Fig.~\ref{fig:dft_fft_inversions}. We compute the relative error, normalized to the peak brightness of the image, for all pixels in the image. We find that the two dirty images match with a maximum error of $1.2 \times 10^{-8}$ and a RMS error of $3.4\times10^{-9}$. Therefore, we conclude that the NUFFT operator produces results that are in excellent agreement with those obtained using the standard DFT-matrix approach.

\subsection{T2: Preconditioner and iterative solution}
\label{sec:T2}

We next test the accuracy of the preconditioned conjugate gradient method (see Section \ref{sec:cgsolver}) when solving equation (\ref{eq:nufftlsq}). Similarly to test T1, we wish to make a comparison with the direct DFT solver and therefore use the small data set $\rm{D_1}$.
We fix the lens parameters to their true values $\etagt$ and the source regularization to the best value inferred from modelling the data with the DFT method, $\lams^\mathrm{DFT,MAP}$. We then
compute the source by inverting equations (\ref{eqn:leastsquares}) and (\ref{eq:nufftlsq}) for the DFT and NUFFT cases, respectively. We compare these reconstructed sources point-by-point, again normalizing to the peak surface brightness. These source reconstructions and the error map are shown in Fig.~\ref{fig:dft_fft_inversions}; we find that the maximum and RMS errors between the source reconstructions obtained with the DFT and NUFFT solvers are $7.6 \times 10^{-7}$ and $1.6 \times 10^{-7}$, respectively.

T2, together with test T1, confirms that the replacement of the DFT matrix with the NUFFT operators and the conjugate gradient solver gives source inversions that are accurate to within a factor of $10^{-6}$. Therefore, for a given $\etalens$ and $\lams$, we consider the NUFFT solver to be equivalent to the DFT implementation within numerical errors. The only remaining potential source of systematic error is in using the log-determinant approximation (see Section \ref{sec:logdet}), which we address in the next section. 

\subsection{T3: Log-determinant approximation}
\label{sec:ldtests}

\begin{figure*}
\centering
\includegraphics[scale=0.55]{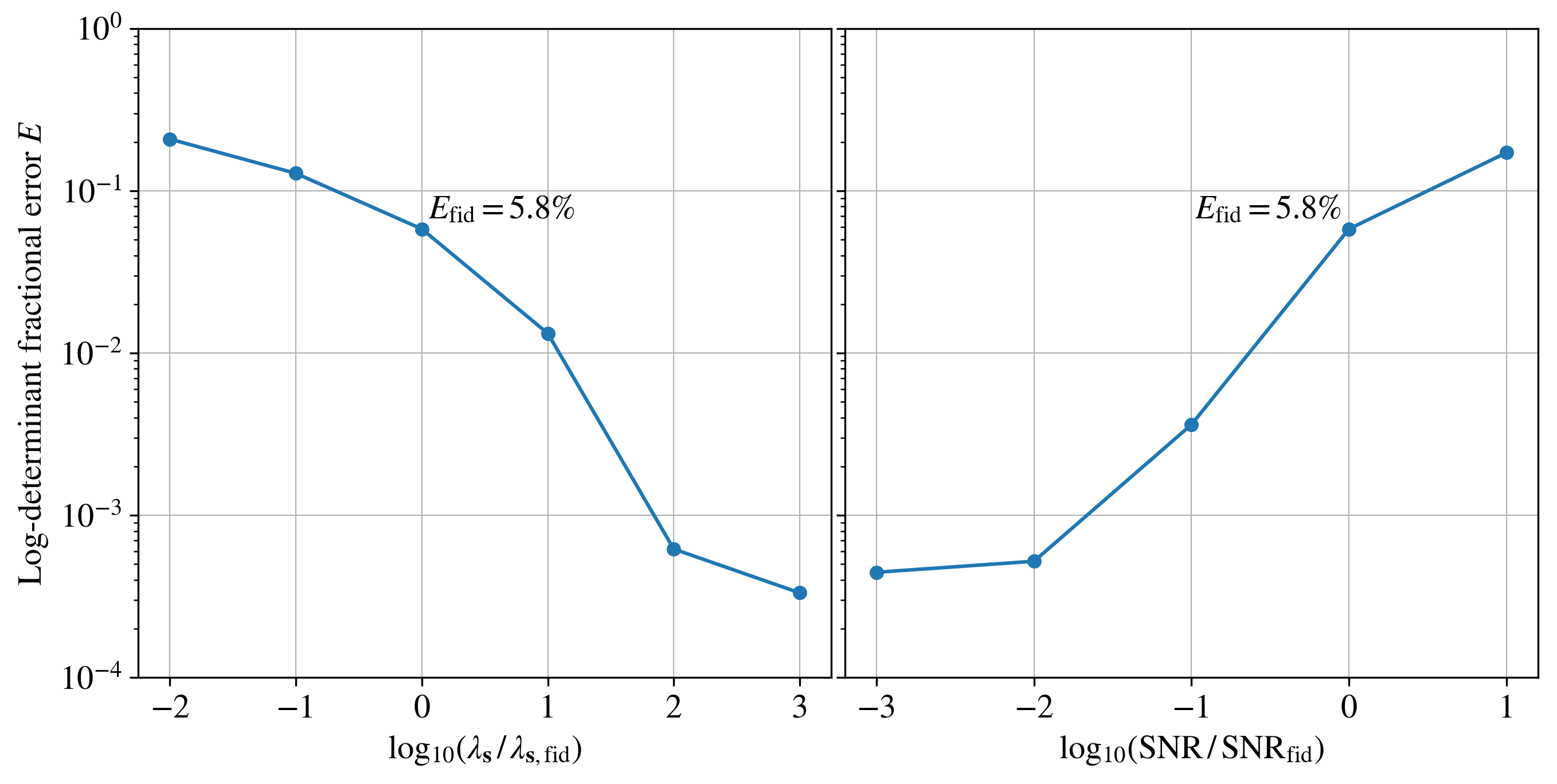}
\caption{The results of test T3. Shown is the fractional error in the approximate log-determinant computation relative to its exact value as computed by the Cholesky decomposition in the DFT implementation, as a function of the source regularization constant (left panel) and the data SNR (right panel).}
\label{fig:logdet}
\end{figure*}

In this section, we examine the accuracy of the approximation to the log-determinant term described in Section \ref{sec:logdet}. 
We begin by fixing the lens parameters to the ground truth, then maximizing the evidence to find the most probable  source regularization for the small data set $\rm D_1$. We refer to this value of $\lams$ as the fiducial value $\lamfid$. We then compare the value of the log-determinant as calculated exactly using the direct solver, and with our numerical approximation, for values of $\lams$ that differ from $\lamfid$ by factors of 10. As shown in the left panel of Fig.~\ref{fig:logdet}, we find that as $\lams$ decreases, the log-determinant approximation worsens to an error of $E \approx 20$~per cent when $\lams = \lamfid \times 10^{-2}$. Conversely, the approximate log-determinant performs well as $\lams$ increases, falling to $E < 0.1$~per cent when $\lams = \lamfid \times 10^{2}$. This behavior arises from the fact that as $\lams$ decreases, the matrix $\msol$ becomes more singular, and hence, more sensitive to its off-diagonal elements.

We conclude that the accuracy of the log-determinant approximation depends on the value of the source regularization, which is in turn tied to the SNR of the data. To fully understand this effect, we turn to the data sets $\rm D_2$ to $\rm D_5$, which have different SNRs that vary in factors of 10 both higher and lower from $\snrfid$. For each of these mock observations, we then optimize $\lams$ to its best value given the SNR. We show the results of this in the right panel of Fig.~\ref{fig:logdet}. We see that for data with an SNR lower than $\sim500$, the log-determinant approximation is accurate to within $1$~per cent due to the stronger prior information (hence higher $\lams$) needed to maximize the evidence. We again find that for an artificially high $\mathrm{SNR}=4.49\times10^4$, the error in the log-determinant is $E \approx 20$~per cent. 

\subsection{T4: Posterior maximisation}
\label{sec:snr}

We next test whether our approximation for the log-determinant is sufficient to accurately infer the lens parameters $\etalens$ given that it exhibits an error of several per cent at our fiducial $\snrfid$ and $\lamfid$. Using the mock data sets $\rm D_1$ to $\rm D_5$ with varying SNR, we optimize for both the best regularization strength $\lams$ and lens parameters $\etalens$. We compare the fractional difference in the resulting lens parameters using our NUFFT approach with those obtained via the DFT implementation and to their ground truth values in Table \ref{tab:dftffteta}. 

\begin{figure*}
 \centering
\includegraphics[scale=0.55]{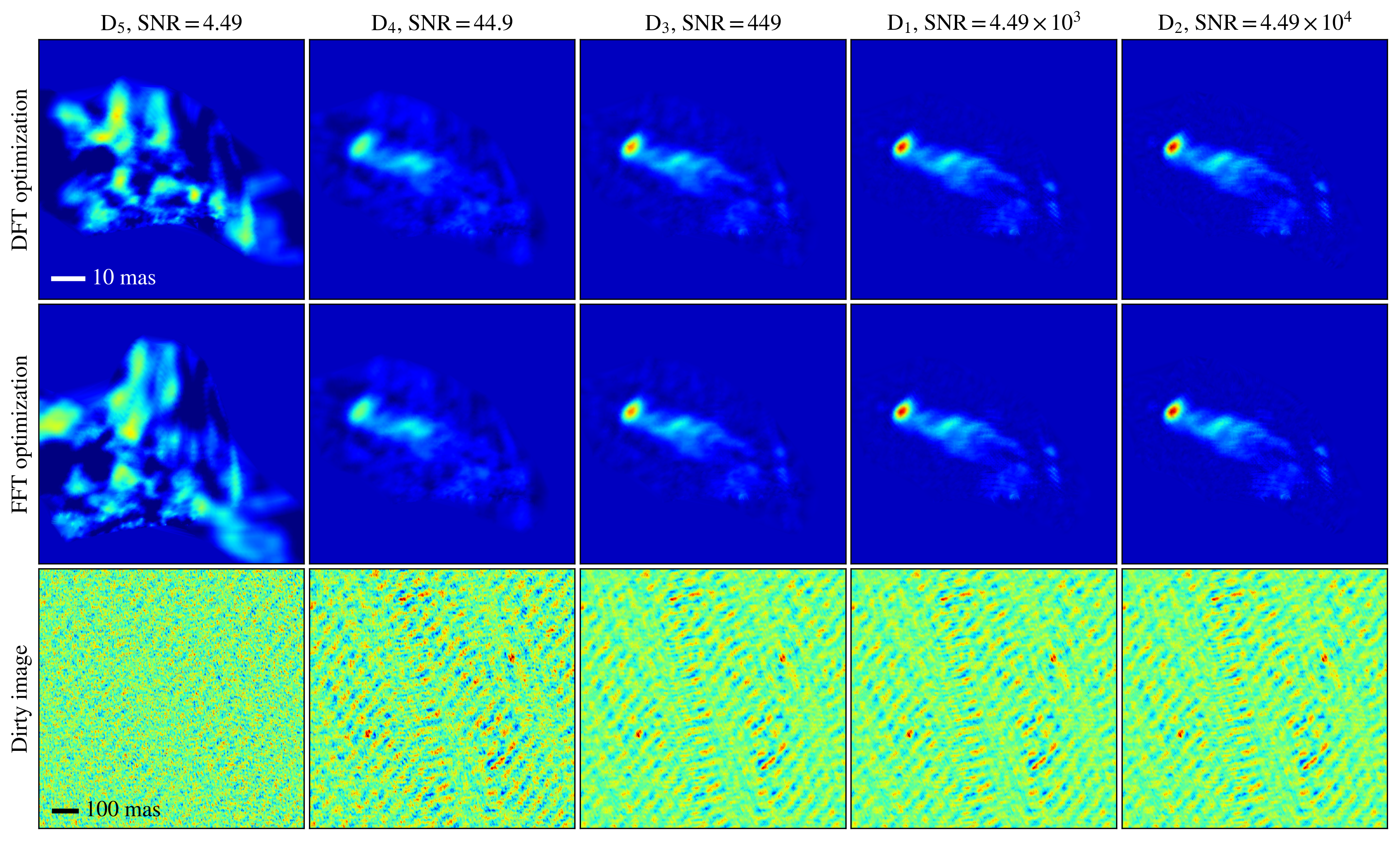}
\caption{The results of test T4. Each panel shows the maximum a posterior source surface brightness distribution inferred by modelling the small data sets $\rm{D_1}$ to $\rm{D_5}$ with the standard DFT approach (top row) and with our new NUFFT modelling technique (\revisions{middle} row). Overall, the recovered source surface brightness distributions using both methods are essentially equivalent. \revisions{Note that for this figure and for Fig.  \ref{fig:opflow}, we have cropped out an emission component to the lower-right of the main jet feature; this is simply to zoom in for more detail and does not reflect any change in the modelling. In the bottom row, we include the dirty image from each data set for a visual comparison between different SNRs.}}
\label{fig:logdetlenspars}
\end{figure*}

Overall, we find that all lens parameters are well recovered to within 1 per cent for all data sets when $\mathrm{SNR} > \snrfid \times 10^{-2}~(\approx 45)$. However, when the SNR is very low ($\sim \snrfid \times 10^{-3}$), the recovery of most lens parameters is poorer, on the order of 1 to 2~per cent, for both the DFT and NUFFT implementations; although, we find that the position $(x_0,y_0)$ and shear strength $\Gamma_\mathrm{sh}$ have much larger fractional errors ($\sim20$~per cent). Furthermore, inspecting the source reconstructions shown in Fig.~\ref{fig:logdetlenspars}, it is clear that the noise is too high for a source to be properly inferred in the case of data set D$_5$ for both the DFT and NUFFT reconstructions. We arrive at the conclusion that, for both the DFT and NUFFT approaches, as long as the SNR in an observation is high enough to discern a source structure, the inference of the lens parameters themselves is robust. Also, we find that the results for the DFT and NUFFT approaches are essentially equivalent.

\subsection{T5: Image-plane mask}
\label{sec:masktest}

Before comparing our reconstructions with the ground truth, we carry out one final test to demonstrate an important point regarding the choice of image-plane masking when modelling interferometric data. Our intent is to show that the choice of mask is itself a form of prior information imposed on the model. To this end, we focus on mock observation $\rm D_6$. 

As mentioned during the description of the lensing operator $\lensop$ in Section \ref{sec:dataresponse}, we do not compute deflection angles for every pixel in the image plane. Rather, only those pixels contained within the mask are used to construct the source model. The implication is that image-plane pixels that lie outside of the mask are assumed to have zero brightness. When modelling optical data, these pixels can simply be excluded from the $\chi^2$ term in the posterior. However, the response of an interferometer is not localized on the sky, meaning that pixels outside of the mask are by necessity included in the model visibilities via the effect of the Fourier transform $\nufft$.

\begin{figure*}
 \centering
  \includegraphics[scale=0.52]{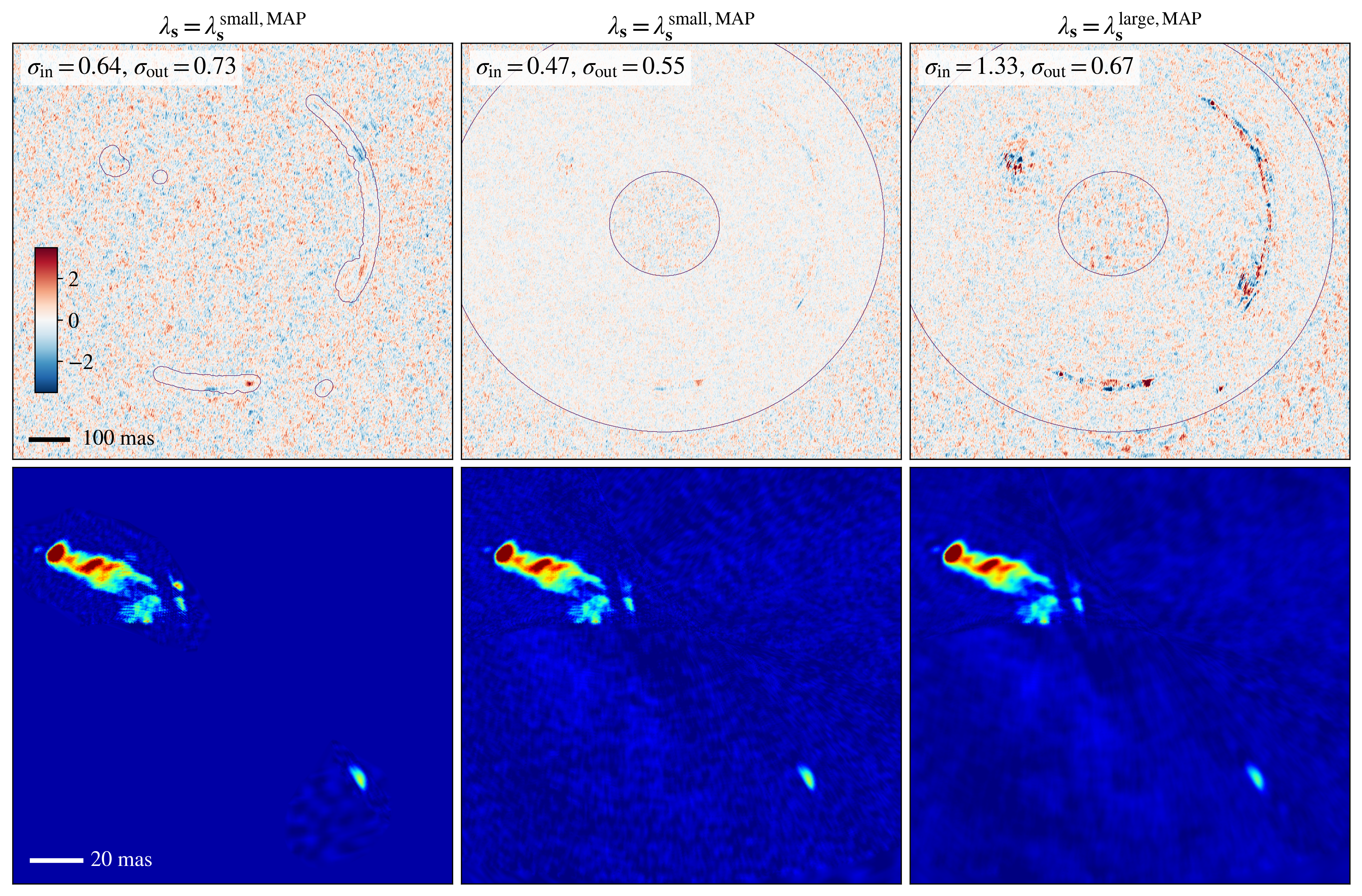}
  \caption{The results of test T5. In the top row, we plot the image plane residuals, calculated as described in Section \ref{sec:masktest}, along with the mask boundaries. In the bottom row we show the corresponding source reconstructions, where we have saturated the colour scale to highlight noise fitting in the source plane. The left column shows the result obtained by modelling the mock data ${\rm D_6}$ using a small mask, optimizing for the best source regularization $\lams^\mathrm{small,MAP}$. In the middle column we show the result of the source inversion using the large mask, but fixing the regularization to the value optimized using the small mask, $\lams^\mathrm{small,MAP}$. In the rightmost column we have modelled the data using the large mask, this time optimizing for $\lams$ to obtain $\lams^\mathrm{large,MAP}$. The inset labels state the residual RMS inside ($\sigma_{\rm in}$) and outside ($\sigma_{\rm out}$) the mask for comparison.}
  \label{fig:masks}
\end{figure*}

In Fig.~\ref{fig:masks}, we show this effect using residual maps obtained by modelling the data with two masks of different sizes.  We create these maps by first normalizing the visibility-space residuals to the noise, then Fourier-transforming into the image plane. We again normalize to $\sqrt{\nvis}$ to give a dimensionless residual map in the image plane. The left and right panels of Fig.~\ref{fig:masks} show the results when the data are modelled with a small and large mask, respectively. In both cases the lens parameters were fixed at the ground truth $\etalens = \etagt$, while the source regularization constant was optimized. 

In the leftmost column of Fig.~\ref{fig:masks}, we show the first example in which we optimize for $\lams$ using the small image-plane mask obtained with the procedure described in Section \ref{sec:mask}. We refer to this source regularization as $\lams^\mathrm{small,MAP}$.  We see that when the mask is tight around the true emission, only image-plane pixels with high SNR are considered by the model.  This allows the optimal regularization strength $\lams^\mathrm{small,MAP}$ to be low and the source structure to be clearly resolved. In the middle panel we plot the result of the source inversion when the large mask is used, but $\lams$ is kept fixed to $\lams^\mathrm{small,MAP}$. Finally, in the rightmost column of Fig.~\ref{fig:masks}, we model the data using the large mask and optimize for the regularization strength to obtain $\lams^\mathrm{large,MAP}$.

\revisions{
We would like to focus the reader's attention on the difference between the centre and left-most columns of Fig.~\ref{fig:masks}. We see that in the center column, in which a large mask is used but the source regularization is fixed to $\lams^\mathrm{small,MAP}$, the source is fit well, giving low residuals throughout the image plane. Some noise features are also fit by the model as well. We contrast this to the rightmost column, in which we re-optimize the source regularization to $\lams^\mathrm{large,MAP}$ using the large mask. Here, the model is more regularized as shown by the smoother residual noise pattern. This comes at the expense of over-regularizing the true source emission, which leads to underfitting around the lensed arcs.  The model prefers different $\lams$ depending on the choice of mask, with $\lams^\mathrm{small,MAP}$ preferring a better fit to the sky emission and $\lams^\mathrm{large,MAP}$ being driven by the presence of additional noise within the mask.
}

\revisions{ 
This effect is due to the interferometer being incapable of spatially distinguishing between signal and noise within the mask.  The regularization $\lams$ is sensitive to the average SNR within the mask, which cannot be localized by the interferometer response. When more noise is present inside the mask, the evidence prefers a smoother source brightness distribution. We can effectively increase the SNR during the reconstruction step by simply masking as much noise as possible out of the image plane.} In other words, the lens modelling process and source reconstruction can be improved by using a mask that is as tight as possible around known emission. Additionally, adding an adaptive regularization, which takes into account the variations in the SNR across the source surface brightness distribution, will help mitigate this issue.

\revisions{``Known emission'' can be difficult to identify robustly in the context of interferometry, due to the challenges inherent in radio imaging ($uv$ coverage, compact vs. diffuse emission, etc.). The creation of a suitable image-plane mask is dependent on a good starting model for the sky emission, which necessarily imposes some prior information on the lens model and source reconstruction. As we describe in Section \ref{sec:mask}, we use a \software{clean}ed image of MG J0751+2716 as our starting model, which assumes nothing about the lens itself. However, in practical settings this approach may fail to identify multiply-imaged dim or diffuse components, which could lead to biases in the source and lens models.  In this paper, we have the benefit of knowing the ground truth sky model exactly, but in future applications to real data this process will require special attention in order to understand the interplay between the choice of mask, the sky emission, the lens model, and the reconstructed source.}

 \subsection{T6: High resolution data from a global interferometric array} 
\label{sec:fullmock}

Here, we model the full mock observation $\rm D_6$ to test our modelling technique in a realistic scenario for high-resolution global VLBI observations of a gravitational lens. As these data are too large to process using a direct matrix solution, motivating this work, we compare our results to the ground truth source after optimizing for both the source regularization constant $\lams$ and the lens parameters $\etalens$.

\begin{figure*}
 \centering
  \includegraphics[scale=0.55]{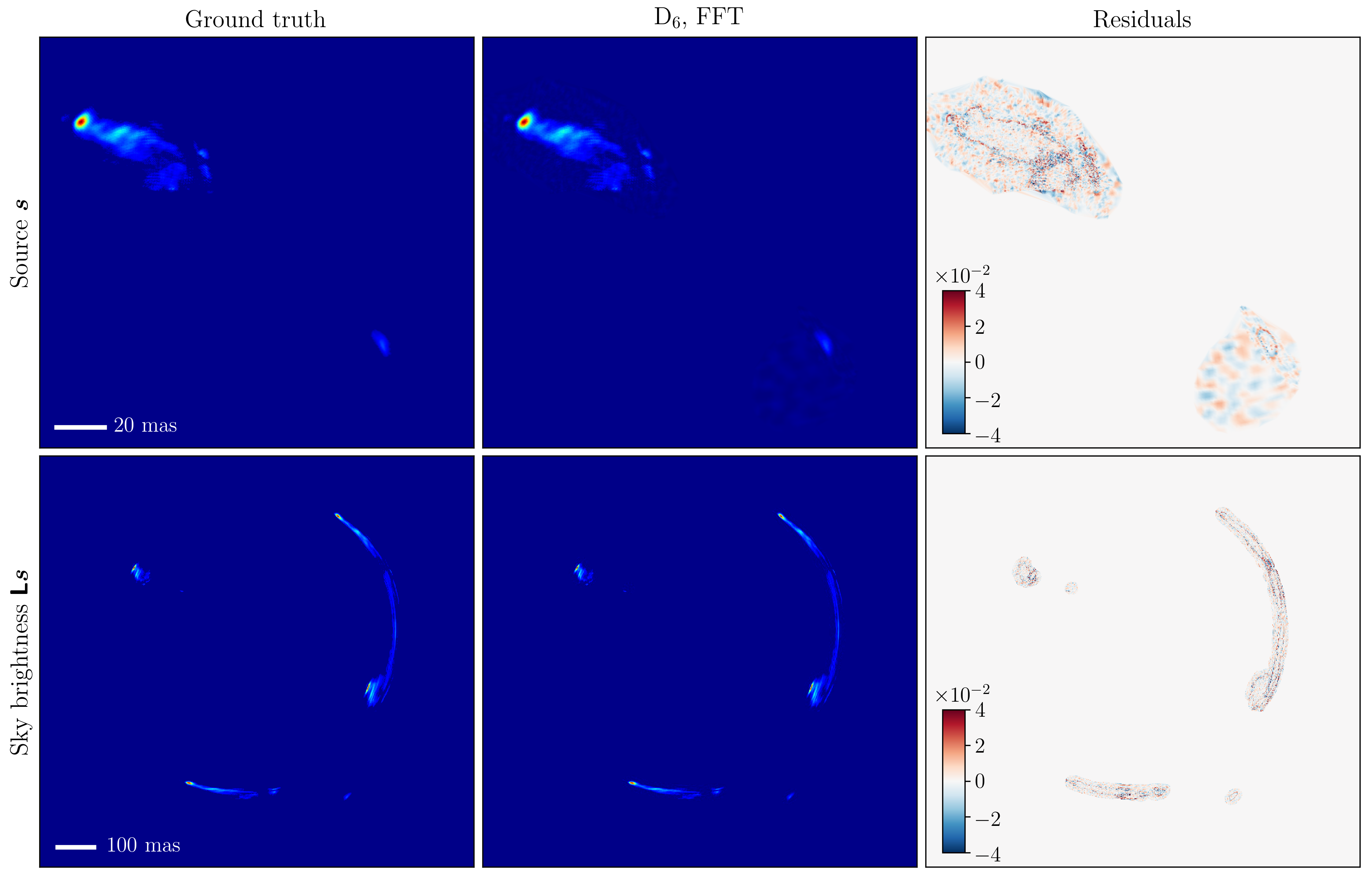}
  \caption{The results for test T6. The top and bottom rows show the source surface brightness and the corresponding sky brightness distributions, respectively. The ground truth, the MAP solution and the difference between the two are displayed in the left, middle and right panels, respectively. The maximum relative residual between the ground truth and MAP solution is 8.3 per cent, while the RMS residual is 0.97 per cent.}
  \label{fig:full_inversions}
\end{figure*}

We show the resulting source reconstruction in Fig.~\ref{fig:full_inversions}. We find the ground truth source and reconstruction to have a maximum relative difference of 8.3~per cent. The RMS difference between the two is 0.97~per cent. However, we do not expect the reconstruction to perfectly recover the ground truth source, as the Bayesian prior imposes additional smoothness constraints on the reconstruction. We also show the reconstructed lensed sky model, which exhibits the same fractional difference relative to the ground truth, due to the conservation of surface brightness by the lensing operator.  We also show the recovered lens parameters in the rightmost column of Table \ref{tab:dftffteta}, which were correctly inferred to within 0.6~per cent in all cases. Overall, for the test case analyzed here, we find that the NUFFT methodology robustly allows the analysis of gravitational lensing data from high resolution interferometric arrays with many collecting elements, which until now was not possible due to the prohibitive size of such data sets.

\begin{table*} 
\centering
\begin{tabular}{c r@{}l r@{}l r@{}l r@{}l r@{}l c }
\hline
Par. & \multicolumn{2}{c}{\makecell{$\mathrm{D_1}$ [\%] \\ DFT~|~GT}} & \multicolumn{2}{c}{\makecell{$\mathrm{D_2}$ [\%] \\ DFT~|~GT}} & \multicolumn{2}{c}{\makecell{$\mathrm{D_3}$ [\%] \\ DFT~|~GT}} & \multicolumn{2}{c}{\makecell{$\mathrm{D_4}$ [\%] \\ DFT~|~GT}} & \multicolumn{2}{c}{\makecell{$\mathrm{D_5}$ [\%] \\ DFT~|~GT}} & \makecell{$\mathrm{D_6}$ [\%] \\ GT} \\
\hline
$\kappa_0$ & $0.0035~$&|~$0.13$ & $0.0012~$&|~$0.057$ & $0.034~$&|~$0.15$ & $0.16~$&|~$0.023$ & $0.085~$&|~$1.00$ & $0.15$ \\
$\gamma$ & $0.0052~$&|~$0.053$ & $0.0036~$&|~$0.021$ & $0.01~$&|~$0.065$ & $0.064~$&|~$0.0066$ & $0.70~$&|~$1.30$ & $0.054$ \\
$q$ & $0.061~$&|~$0.031$ & $0.13~$&|~$0.014$ & $0.047~$&|~$0.095$ & $0.022~$&|~$0.76$ & $0.76~$&|~$0.77$ & $0.03$ \\
$\theta$ & $0.035~$&|~$0.021$ & $0.28~$&|~$0.19$ & $0.23~$&|~$0.066$ & $0.69~$&|~$0.85$ & $1.0~$&|~$1.7$ & $0.33$ \\
$x_0$ & $0.00~$&|~$0.0094$ & $0.0081~$&|~$0.0083$ & $0.0018~$&|~$0.022$ & $0.079~$&|~$0.13$ & $0.42~$&|~$1.10$ & $0.0063$ \\
$y_0$ & $0.013~$&|~$0.028$ & $0.035~$&|~$0.022$ & $0.0073~$&|~$0.066$ & $0.044~$&|~$0.067$ & $7.4~$&|~$6.5$ & $0.04$ \\
$\Gamma_\mathrm{sh}$ & $0.05~$&|~$0.11$ & $0.019~$&|~$0.29$ & $0.11~$&|~$0.081$ & $0.25~$&|~$1.10$ & $14~$&|~$24$ & $0.59$ \\
$\theta_\mathrm{sh}$ & $0.043~$&|~$0.026$ & $0.22~$&|~$0.026$ & $0.097~$&|~$0.0071$ & $0.69~$&|~$0.057$ & $0.38~$&|~$0.33$ & $0.069$ \\
\hline
\end{tabular}
 \caption{The results of tests T4 and T6, in terms of the recovered lens mass parameters. For each parameter, we quote the relative error with respect to the results obtained with the DFT direct solution approach (when possible) and the ground truth GT.}
\label{tab:dftffteta}
\end{table*}

\section{Conclusions and Future Prospects}
\label{sec:conclusions}

In this paper, we have presented a novel approach for modelling high angular-resolution strong gravitational lensing observations from many-element interferometers, such as from a global VLBI array, directly in the $uv$ plane. Previous approaches to lens modelling in visibility space have been challenged by the large numbers of visibilities typical of current interferometric arrays. As such, they have relied on averaging in time and frequency or otherwise reducing the size of the data in preprocessing, which can adversely affect the data quality and limit the scientific interpretation. Observations at high angular resolution also require large numbers of image-plane pixels, which compounds the computational demands of the modelling process. Improving on existing techniques, our method introduces numerical solutions that overcome previous limitations on the size of the data arising from the prohibitive memory and computing time required by a direct matrix solver. We have incorporated them into a Bayesian lens modelling framework capable of jointly inferring the source brightness distribution and lens mass model directly from data sets with milli-arcsec angular resolution, with no prior averaging required. Using realistic mock observations, we have thoroughly tested our method for varying levels of SNR and different combinations of free parameters.  We have compared its  performance with respect to the ground truth as well as the more standard visibility-fitting approach based on a direct matrix solver, when computationally feasibile. Our results can be summarised as follows.

\begin{enumerate}
    \item \label{item1} Replacing the DFT operator with an NUFFT has a negligible effect on the calculation of the dirty image from the data, which is a key ingredient in the calculation of the most probable a posteriori source.
    \item Likewise, replacing the dense image-plane covariance matrix $\imcovexplicit$ with a FFT-based dirty beam convolution also has a negligible effect on the MAP source inversion. We find that when iteratively solving equation \eqref{eq:nufftlsq} using this fast image-plane covariance operator $\imcov$ and the NUFFT operator $\nufft$, we obtain a source reconstruction with a maximum error of no more than $10^{-6}$ relative to the direct DFT solution.
    \item We introduce a sparse approximation to the MAP source inversion matrix $\msol$, which forms a preconditioner for the conjugate gradient solver. This drastically improves the convergence rate of the solution.
    \item We approximate $\logdet \msol$ using the log-determinant of the preconditioner matrix, which is obtained at no additional cost when setting up the iterative solver. We find that for a fiducial mock observation mimicking the SNR of the true  MG J0751+2716 observation, this approximation is accurate to within 6~per cent. This approximate log-determinant becomes more accurate when modelling observations with a lower SNR, as they result in a larger value of the source regularisation level. However, in practice, the approximate log-determinant does not affect our ability to infer correctly the most probable lens parameters $\etalens$ and source regularization level. 
    \item As long as the SNR is sufficiently high ($\mathrm{SNR} > 5$) to discern the source structure, we can recover the lens parameters and the source to within 0.6 per cent. Below this threshold both the DFT and our NUFFT approach result in larger errors, indicating that this is related to the \revisions{SNR of the data and choice of statistical priors rather than the modelling technique itself}.
    \item We examine the effect of the image-plane mask on overfitting. We determine that a mask that is as tight as possible around the known emission is best for preventing overfitting to the noise, since the source regularization is not dominated by suppressing the noise within the mask. This is due to the non-localized response of the interferometer to the sky brightness, which can potentially confuse noise for sky emission in unmasked regions.
    \item We finally run our full modelling pipeline on a full mock data set that exactly mirrors the true global VLBI observation of MG J0751+2716 in $uv$ converage, number of visibilities, and noise.  We recover a source with RMS surface brightness errors at 0.97~per cent and lens parameters accurate to fractions of a per cent.
\end{enumerate}

Our improved approach to the visibility-space modelling of strong gravitational lens data provides a competitive alternative to the traditional method based on the use of a direct matrix solver. Our solution produces results of equal quality with the advantage of not being limited by memory, speed, or the need to average the data in time or frequency. This clears the path towards modelling high-resolution data sets with a large number of visibilities without loss of information. The tests that we have carried out have focused on applications to very high angular-resolution global VLBI observations at cm-wavelengths. However, our methodology can also be applied to gravitational lensing data from other interferometers, for example, the Low Frequency Array (LOFAR), the Jansky Very Large Array (VLA), the Multi-Element Remotely Linked Interferometer Network (e-MERLIN) and ALMA. Each of these arrays provide their own set of analysis challenges due to their large number of antennas, large bandwidths, or high angular-resolutions that in principle can be overcome with our new modelling approach. In the future, the Square Kilometre Array (SKA) will have a large number of collecting elements ($\sim 200$ antennas), and methods such as those presented here will be important for gravitational lensing studies with this next generation radio telescope.

Our tests have also assumed a perfectly calibrated data set that does not have any residual amplitude and phase errors on the visibilities. Self-calibration techniques have been developed and extensively used for interferometric data sets to reduce calibration errors, which can be applied during the pre-processing steps before lens modelling (as was done here). However, incorporating self-calibration in a self-consistent way into our methodology (e.g. see \citealt{hezaveh2013,arras2019}) can in principle provide a more robust calibration of the data. This is because the sky surface brightness distribution will be correlated due to the lens modelling. This is a topic that will be the focus of a follow-up publication.

In addition, we have assumed a lens model that is perfectly described by an elliptical power-law mass distribution, which is likely insufficient to describe the structure of galaxies on small-scales. Indeed, an initial analysis of the global VLBI observations of MG J0751+2716 by \citet{spingola2018} found that the image positions could not be reproduced by such a model, with an rms of $\sim3$~mas between the observed and predicted positions. This has implications for both recovering a robust source reconstruction from high angular-resolution data, or in detecting low mass structure within the lens or along the line of sight to test models for dark matter. Next, we plan to incorporate pixellated potential corrections to the lensing mass model (e.g. see \citealt{vegetti2009}) to account for the additional mass complexity within the lens, which will allow a better estimate of the surface brightness distribution of a high redshift source on pc-scales and an investigation of the halo mass function in the mass regime of $\sim 10^6$~M$_\odot$

\section*{Acknowledgements}
\revisions{We thank Philipp Arras, Matt Auger, Torsten Ensslin, Chris Fassnacht, Leon Koopmans, Andr\'e Offringa, and the referee Yashar Hezaveh for useful comments and insightful discussions. We thank Matus Rybak for early contributions to the project.}  We also would like to express our gratitude to the \software{petsc} support team at Argonne National Lab: Satish Balay, Matthew Knepley, Karl Rupp, Junchao Zhang, Jed Brown, and Barry Smith. SV has received funding from the European Research Council (ERC) under the European Union's Horizon 2020 research and innovation programme (grant agreement No 758853). JPM acknowledges support from the Netherlands Organization for Scientific Research (NWO) (Project No. 629.001.023) and the Chinese Academy of Sciences (CAS) (Project No. 114A11KYSB20170054). CS is grateful for support from the National Research Council of Science and Technology \markup{(NST)}, Korea (EU-16-001). The National Radio Astronomy Observatory is a facility of the National Science Foundation operated under cooperative agreement by Associated Universities, Inc. The European VLBI Network \markup{(EVN)} is a joint facility of European, Chinese, South African and other radio astronomy institutes funded by their national research councils. Scientific results from data presented in this publication are derived from the following EVN project code: GM070.

\revisions{
\section*{Data availability}
The data underlying this article will be shared on reasonable request to the corresponding author.
}

\bibliographystyle{mnras}
\bibliography{j0751_methods} 



\appendix

\section{The Kaiser-Bessel gridding kernel} \label{app:kb}

We briefly show the Kaiser-Bessel gridding kernel, its inverse Fourier transform, and the optimal shape parameter. This is a re-statement of Section IV from \cite{beatty2005}, with slightly different notation. The Kaiser-Bessel kernel is separable (it is a product of independent functions in the $x$ and $y$ directions), so we work in one dimension here.

The gridding kernel itself is
\begin{equation}
 K_k(k) = 
\begin{dcases}
   \frac{1}{2\wsup} I_0 \left( \beta \sqrt{ 1 - \left( \frac{k}{\wsup} \right)^2  } \right), & |k| \leq \wsup \\ 
    0, & |k| > \wsup\,.
\end{dcases}
\end{equation}
Here, we work in dimensionless units where $k$ is a number of gridpoints in the Fourier plane. We also note that $\wsup$ is the support \emph{radius}, in contrast to the notation of \cite{beatty2005}, who state the kernel support in terms of its diameter. $I_0$ is the zeroth-order modified Bessel function of the first kind.

The inverse Fourier transform of the gridding kernel is 
\begin{equation}
   K_x(x) = \frac{\sinh{\left( \sqrt{ \beta^2 - (2 \pi \wsup x)^2} \right)}}{\sqrt{ \beta^2 - (2 \pi \wsup x)^2}}\,,
\end{equation}
where the dimensionless coordinate $x$ is the number of grid points in the image plane. This inverse Fourier transform forms the apodization correction operator $\apod$ (see Section \ref{sec:nufft}), which de-convolves the effect of the gridding kernel from the final image.

The parameter $\beta$ is a shape parameter that can be chosen to minimize the image reconstruction error given a support radius $\wsup$ and oversampling ratio $\alpha$. \cite{beatty2005} give  an expression for choosing the shape parameter in a simple, but near-optimal way:
\begin{equation}
    \beta = \pi \sqrt{ \frac{4\wsup^2}{\alpha^2} \left(\alpha-\frac{1}{2}\right)^2 - 0.8 }\,.
\end{equation}
For our $\alpha=2$ and $\wsup=4$, we therefore adopt $\beta = 18.64$ for this work.

\section{Derivation of the image-plane covariance} \label{app:dbder}

Here, we derive equation \eqref{EQ:DBCONV}, the image-plane noise covariance $\imcovexplicit = \dft^T \noisecov \dft$.  In the numerical implementation of this method, we work entirely in real units where the data vector contains separate rows for its real and imaginary components. In this derivation, however, we write in complex quantities for ease of notation.  

The non-uniform direct Fourier transform is 
\begin{equation}
    \dft_{kj} =  \exp \left ( - 2 \pi \mathcal{I} \uu_k \cdot \xx_j   \right)\,,
\end{equation}
where $\uu_k$ is the $uv$ coordinate of the $k$\textsuperscript{th} visibility, and $\xx_j$ is the position of the $j$\textsuperscript{th} pixel in the image plane. We write the imaginary unit as $\mathcal{I}$ to avoid confusion with the index $i$. This matrix maps the image-plane surface brightness onto Fourier modes observed by the interferometer. 
Since $\dft$ is a complex operator, its transpose includes a complex conjugation ($\dft_{kj}^T = \dft_{jk}^*$). 

The noise covariance in visibility space is
\begin{equation}
    \noisecov_{kl} = \delta_{kl} \frac{1}{\sigma_k^2}\,,
\end{equation}
where we will see in a moment that $\sigma_k$ need only be the noise associated with the real part of visibility $k$. It is then clear via the index notation that combining the three operators gives
\begin{align} 
    \left( \nufft^T \noisecov \nufft  \right)_{ij} &= \exp \left (2 \pi \mathcal{I} \uu_l \cdot \xx_i   \right) \,\left ( \delta_{kl} \frac{1}{\sigma_k^2}\right) \, \exp \left (- 2 \pi \mathcal{I} \uu_k \cdot \xx_j   \right) \\
    &= \sum_{k} \,  \frac{1}{\sigma_k^2} \exp \left (- 2 \pi \mathcal{I} \uu_k \cdot (\xx_j - \xx_i)  \right)\,.
\end{align}

The final step involves recognizing that the visibility data is Hermitian; that is, for every visibility $k$, there exists a conjugate visibility on the $uv$ plane corresponding to the exchange of the two antennas used to measure it. For a visibility $\data_k = d(\uu_k)$, this implies the existence of the conjugate visibility $d^*(\uu_k) = d(-\uu_k)$.  We exploit this fact by applying the identity $\left[\exp(\mathcal{I}\theta) + \exp(-\mathcal{I}\theta) \right]/2 = \cos(\theta)$ to arrive at the result
\begin{equation} 
    \oper{C}_{x,ij}^{-1} = \sum_{k} \, \frac{1}{\sigma_k^2} \cos\left[2\pi \uu_k \cdot (\xx_i-\xx_j) \right]\,.
\end{equation}

 \section{Speed} 
\label{sec:performance}

\begin{table*} 
\centering
\begin{tabular}{c c c c c c c c}
\hline
Data Set & $\nvis$ & $\nsource$ & Method & Operator construction (s) & Dirty image (s) &  Linear solution (s)  &  Evidence computation (s)  \\
\hline
$\rm D_1$ & $2.8\times10^3$ & $1.0\times10^4$ & DFT & 35.1 & 0.04 & 1030 & 0.13  \\
$\rm D_1$ & $2.8\times10^3$ & $1.0\times10^4$ & FFT & 2.80 & 0.51 & 168 & 0.41  \\
$\rm D_6$ & $2.4\times10^8$ & $3.2\times10^4$ & FFT & 197 & 101 & 132 & 92 \\
\hline
\end{tabular}
 \caption{\revisions{The results of test T7, showing sample wall times for various execution stages of our modelling code. These tests were run using four MPI tasks and (in the FFT case) two GPUs. The performance will of course vary depending on the specific hardware configuration; these numbers are intended only to give a general sense for execution time and comparison between the DFT and FFT-based methods for different data sizes and source-plane resolutions. We describe this test in detail in Section \ref{sec:performance}. }}
\label{tab:timing}
\end{table*}

\revisions{
Lastly, we briefly present the results of some performance tests in Table \ref{tab:timing}.  These tests were run on four processors and (in the FFT case) two Nvidia Tesla GPUs; this choice was informed by the availability of two GPUs per node which had to be shared between MPI tasks.  These reported times will of course vary depending on the harware used.  Therefore, this section is not meant to be a detailed performance profile, but rather to give a general sense of how long each step of the modelling process takes depending on the data size and source-plane resolution.
}

\revisions{
 We split the times into four categories.  First is the operator assembly, which consists of either populating the full DFT matrix or creating the operators comprising the NUFFT (dirty beam convolution, zero-padding, etc.). Second is the creation of the dirty image, which is needed for the right-hand side of eqn. \eqref{eqn:leastsquares}.  Third is the solution step itself. We include the assembly of the lensing operator and the regularization matrix in this step. For the DFT case, we also include the assembly of the solution matrix, the Cholesky decomposition, and the back-substitution in the reported solution time. For the FFT-based solver, we include the assembly and factorization of the preconditioner, and the conjugate gradient solution. Last is the evidence evaluation, which requires projecting the MAP reconstructed source back into visibility space before forming the $\chi^2$.  The solution and evidence evaluation must be repeated while optimizing for the non-linear lens parameters $\etalens$ until convergence is met; we typically find this to be $\sim100$ iterations for the problems in this paper.   }
 
 \revisions{
 It is clear to see that, even with the highly reduced $\nvis$ and $\nsource$ used for the tests on data set $\rm D_1$, the solution step using the DFT is $\sim 6$ times slower than that of the FFT-based method. This is due to both the sheer number of floating point operations needed to perform the dense matrix arithmetic when assembling the solution matrix, as well as the poor scaling of direct linear solvers, which go as $\mathcal{O}(N^3)$ in the problem dimension. This test is, in fact, the largest problem size that we could reasonably run the DFT solver on for the tests in this paper.
 }
 
 \revisions{
 The FFT-based solutions show much better scaling with the problem size. We see that the strongest dependence is in the operator assembly, the dirty image computation, and the evidence evaluation.  This is due to the fact that each of these steps requires a gridding or de-gridding operation, which scales as $\nvis$. It also illustrates the importance of GPU acceleration in our method.}
 
 \revisions{The solution steps in the FFT-based solver are roughly equal for both the test using $\rm D_1$ and $\rm D_6$. This is because the solution step itself scales with $\nsource$ rather than $\nvis$: In this solution method, all of the operations involving the gridding operation (creating the dirty beam convolution operator, for example; see Section \ref{sec:imcov}) can be carried out ahead of time, so no $\mathcal{O}(\nvis)$ operations enter into the conjugate gradient solver itself.  For the FFT-based tests in this paper, we observe that after optimizing for $\lams$, the CG solve requires a few hundred iterations to achieve convergence. This number depends on both the convergence tolerance and on the characteristics of the data (SNR, $uv$ coverage, etc.) which help set the condition number of the problem. For this reason, the larger data set $D_6$ actually requires slightly fewer iterations to reach convergence, as reflected by the times given in the bottom two rows of Table \ref{tab:timing}.  This, however, is part of a more technical discussion on linear algebra and iterative solvers which lies outside the scope of this paper.
}


\bsp	
\label{lastpage}
\end{document}